\documentclass[%
 reprint,
 amsmath,amssymb,
 aps,
prb,
]{revtex4-2}
\usepackage{amsmath}
\usepackage{graphicx}
\usepackage{dcolumn}
\usepackage{bm}
\usepackage{amsmath}
\usepackage{lineno}
 \usepackage{hyperref}
 \hypersetup{
     colorlinks=true,
     linkcolor=blue,
     filecolor=blue,
     citecolor = blue,      
     urlcolor=blue,
     }

\usepackage{tikz,xcolor,hyperref}

\definecolor{lime}{HTML}{A6CE39}
\DeclareRobustCommand{\orcidicon}{%
	\begin{tikzpicture}
	\draw[lime, fill=lime] (0,0) 
	circle [radius=0.16] 
	node[white] {{\fontfamily{qag}\selectfont \tiny ID}};
	\draw[white, fill=white] (-0.0625,0.095) 
	circle [radius=0.007];
	\end{tikzpicture}
	\hspace{-3.5mm}
}

\foreach \x in {A, ..., Z}{%
	\expandafter\xdef\csname orcid\x\endcsname{\noexpand\href{https://orcid.org/\csname orcidauthor\x\endcsname}{\noexpand\orcidicon}}
}

\begin{document}

\preprint{APS/123-QED}

\title{Polaritonic Critical Coupling in a Hybrid Quasi-Bound States in the Continuum Cavity-WS$_2$ Monolayer System}
\author{Xia Zhang\orcidA{}}
\email{xzhang@tcd.ie}
\author{A. Louise Bradley\orcidB{}}%
 \email{bradlel@tcd.ie}
\affiliation{%
School of Physics, CRANN and AMBER, Trinity College Dublin, Dublin, Ireland
}%

             

\begin{abstract}
We theoretically propose and numerically demonstrate that perfect feeding of a polaritonic system with full electromagnetic energy under one-port beam incidence, referred to as polaritonic critical coupling, can be achieved in a hybrid dielectric metasurface-WS$_2$ monolayer structure. Polaritonic critical coupling, where the critical coupling and strong coupling are simultaneously attained, is determined by the relative damping rates of the cavity resonance, $\rm \gamma_Q$, provided by a symmetry-protected quasi-bound states in the continuum, and excitonic resonance of WS$_2$ monolayer, $\rm \gamma_X$. We reveal that the population of the polariton states can be tuned by the asymmetric parameter of the quasi-bound states in the continuum. Furthermore, polaritonic critical coupling is achieved in the designed system while $\rm \gamma_Q=\gamma_X$ and only strong coupling is achieved while $\rm \gamma_Q\neq\gamma_X$. This work enriches the study of polaritonic physics with controlled absorbance and may guide the design and application of efficient polariton-based light-emitting or lasing devices.

\end{abstract}

\maketitle


\section{Introduction}

A photon emitter placed in an optical cavity interacts with the cavity and experiences a change in the photonic density of states. When the interaction rate is slower than their average incoherent dissipation rates, the system operates in the weak coupling regime \cite{PhysRevLett.97.017402, novotny2011antennas,torma2014strong,curto2010unidirectional}. However, when the coherent rate dominates, half-light, half-matter bosonic quasiparticles are formed, termed as polaritons \cite{PhysRevB.77.115403,PhysRevLett.93.036404, PhysRevLett.118.073604,chikkaraddy2016single, peter2005exciton,PhysRevLett.93.036404}. The system operates in the strong coupling or polaritonic coupling. Whatever the regime, weak or strong coupling, maximizing the absorbance is a fundamental property of light-matter interaction and is critical to a wide range of applications, such as photoluminescence enhancement \cite{wang2016giant, PhysRevLett.97.017402}, nonlinear harmonic generation \cite{michaeli2017nonlinear,koshelev2020subwavelength}, lasing \cite{Bhattacharya2014,de2009stimulated,wu2015monolayer} and quantum correlations \cite{van2013photon, munoz2019emergence}. The dissipation of all the electromagnetic energy fed into the system within the system itself corresponds to perfect absorption which occurs when critical coupling is achieved \cite{haus1984waves}. The underpinning physics of critical coupling is impedance matching \cite{PhysRevLett.100.207402, Radi2015}, or the balance of the radiative rate (scattering) with the intrinsic loss rate (dephasing or absorption) of the hybrid system \cite{Tormo2015, piper2014total,Xiao2020}, which does not rely on the coupling strength of the cavity and the emitter. Coherent perfect absorption can be achieved in a single cavity owning coupled resonances \cite{gorodetsky1999optical,cai2000observation,Tischler:06} or in a hybrid cavity-emitter system, supporting the resonant cavity mode and emitter's excitonic resonance \cite{noh2012perfect,zhu2016coherent}. 

To date, critical coupling has been realized and more documented in the weak coupling regime with a view to tailoring the absorbance bandwidth or magnitude \cite{piper2014total,Epstein2020,Xiao2020}. However, within the strong coupling regime, most reports focus on theoretical or experimental demonstration of the generation of exciton-polariton states \cite{bellessa2004strong, Liu2014crystal, chikkaraddy2016single, Liu2017,Xie2020,cao2020normal, lawless2020influence}, with only very few reports exploring the absorbance magnitude of polariton states \cite{zanotto2014perfect,Baldacci2015,Li2017critical,Qin2021}.  Feeding or pumping the polariton system with maximum electromagnetic energy is vital for the efficiency of polariton-based devices \cite{Sanvitto2016}. A pioneering work in Ref. \cite{zanotto2014perfect} shows that critical coupling and strong coupling can be simultaneously achieved in one system with maximum absorbance by manipulating the coherent rate and damping rate, termed as polaritonic critical coupling. Rather than trying to minimise the damping rate to aim for a high-$\rm Q$ cavity, we explore exploiting and tuning the cavity damping rate to achieve critical coupling in the strong coupling regime.

A dielectric metasurface cavity and two-dimensional monolayer of WS$_2$ are used as the photonic and excitonic resonators, respectively. Generally, dielectric resonators display an advantage over the plasmonic counterpart due to low Ohmic losses \cite{Staude2017,hugall2018plasmonic}. More specifically, a dielectric resonant cavity employing quasi-bound states in the continuum (Q-BIC) is chosen as it can control the damping rate through the structure's asymmetry parameter \cite{koshelev2018asymmetric, koshelev2018strong,wang2020controlling}, therefore, Q-BIC structure is ideal for probing the effect of the damping rate in the light-matter interaction. The Q-BIC structure has displayed extraordinary spatiotemporal field confinement \cite{zhang2021ultra}. Moreover, the Q-BIC cavity has negligible absorbance loss, and the multipolar modes' radiative scattering is dominant. Also in sharp contrast with plasmonic structure employing mainly the electric dipole resonance, magnetic dipole and higher order quadrupole resonances also play a role, where the coherent interplay of all multipolar modes offers peculiar scattering patterns \cite{evlyukhin2016optical} and thus provides the possibility for inspecting some particular modes for coupling.  WS$_2$ monolayer is chosen as the two-level atomic emitter due to its direct band-gap, high in-plane transition dipole moments, optical stability and atomic thickness \cite{PhysRevLett.105.136805, li2014measurement,Epstein2020}. The paper is organized as follows. Firstly within the temporal coupled-mode-theory, the conditions for critical coupling, strong coupling as well as polaritonic critical coupling are explored. Secondly, a proof-of-concept demonstration that critical coupling and polaritonic critical coupling can be separately achieved in the hybrid Q-BIC cavity with monolayer WS$_2$ system is presented. The system is tuned between the different regimes only by varying the damping rate of the Q-BIC cavity. The contributions of the different multipolar modes within the resonator are also revealed. 

\begin{figure*}[htbp]
\includegraphics[width=1\linewidth]{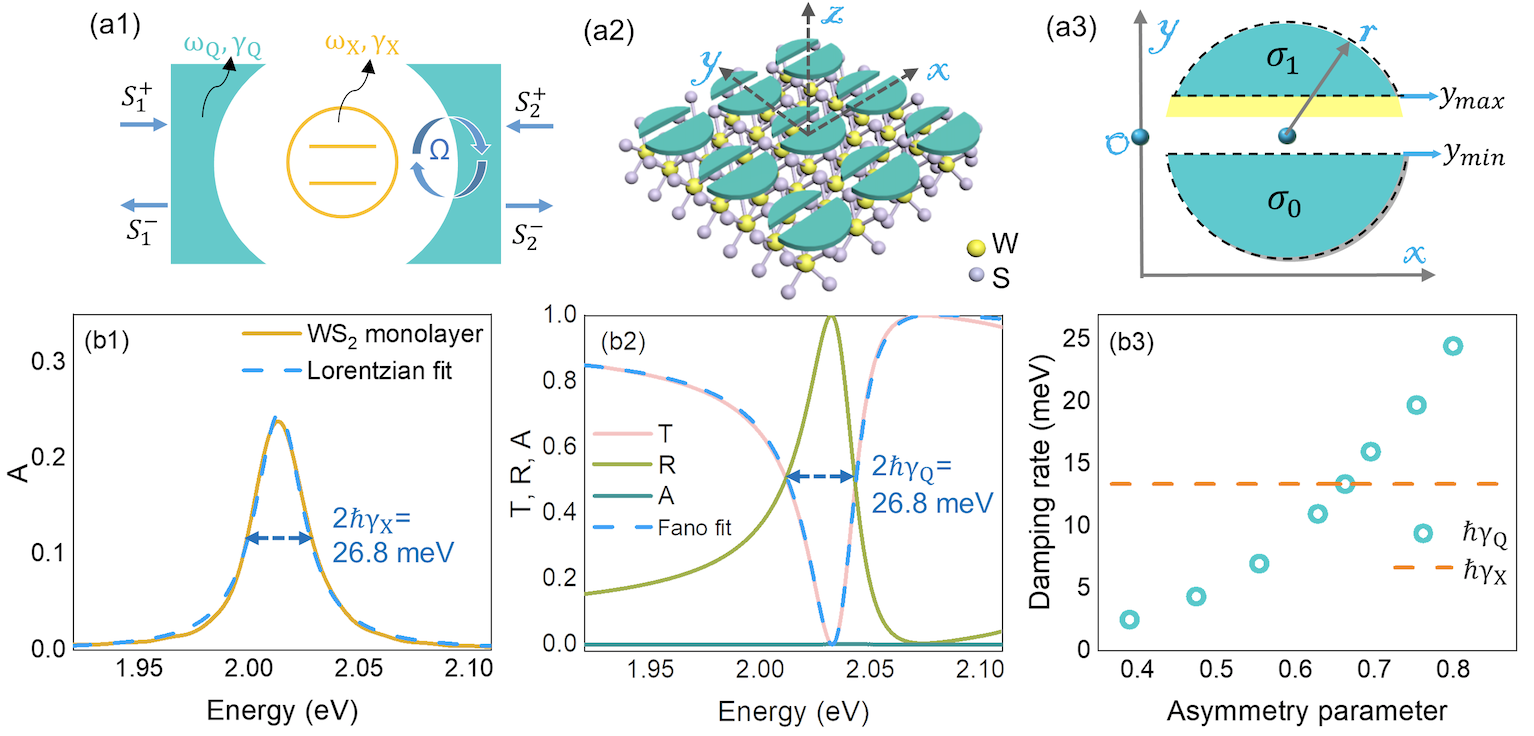}
\caption{\label{fig:sch} (a1) Schematic of the coupled cavity-emitter system. A cavity with the damping rate, $\rm \gamma_Q$ and resonant frequency, $\rm \omega_Q$, is coupled to an emitter, with the damping rate, $\rm \gamma_X$ and resonant frequency, $\rm \omega_X$. $\rm \Omega$ is the coupling constant between the emitter and cavity resonators. The hybrid cavity-emitter system can be coupled with $m$ ports, with the example of $m$ = 2 shown in the schematic. $\rm S^+$, $\rm S^-$ represent the amplitudes of the incoming and outgoing waves at each port. (a2) Schematic of the Q-BIC cavity-emitter system, which includes monolayer WS$_2$ as an 
exciton emitter and a metasurface Q-BIC gap cavity. The incident beam propagates along $z$ and is polarized along $x$ as $\rm \textbf{E}_{inc}=E_0e^{ik_0 z-\omega t}\textbf{x}$. (a3) The unit cell of the Q-BIC gap cavity metasurface. $\rm y_{min}$ and $\rm y_{max}$ denote the position of the gap, which is symmetric while $\rm y_{max}=y_{min}$, otherwise symmetry-breaking exists. $\rm r$ refers to the radius of the disk. The yellow shaded area indicates the reduced disk area $\rm \sigma_1$ relative to $\rm \sigma_0$ due to symmetry-breaking. $\rm \sigma_0$ and $\rm \sigma_1$ are illustrated as circled areas by black dash lines. The asymmetry parameter is calculated as $\rm \alpha=1-\sigma_1/\sigma_0$. (b1) Simulated single-beam absorbance spectra of the WS$_2$ monolayer in air and a Lorentzian fit of the spectrum, from which the spectral width of the exciton resonance, corresponding to the damping rate is extracted, $\rm 2\hbar\gamma_X$ = 26.8 meV.  (b2) Single-beam reflectance (R), transmittance (T) and absorbance (A) spectra of the lossless Q-BIC metasurface in air, where the period $\rm p_x=p_y$ = 550 nm, $\rm y_{min}$ = - 20 nm, $\rm y_{max}$ = 144 nm, radius $\rm r$ = 240 nm, height $\rm h$ = 60 nm. A Fano fit of the transmittance spectrum yields a damping 
rate of $\rm 2\hbar\gamma_Q$ = 26.8 meV. (b3) The damping energy of the Q-BIC metasurface, $\rm \gamma_Q$ versus asymmetric parameter $\rm \alpha$, compared with that of WS$_2$ monolayer, $\rm \gamma_X$. The damping rates are changed by tuning the asymmetry parameter of the gap cavity, while maintaining the radius of the disk, $\rm r$ = 240 nm, the height of the disk $\rm h$ = 60 nm and $\rm y_{min}$ = - 20 nm fixed. }
\end{figure*}

\begin{figure*}[htbp]
\includegraphics[width=1\linewidth]{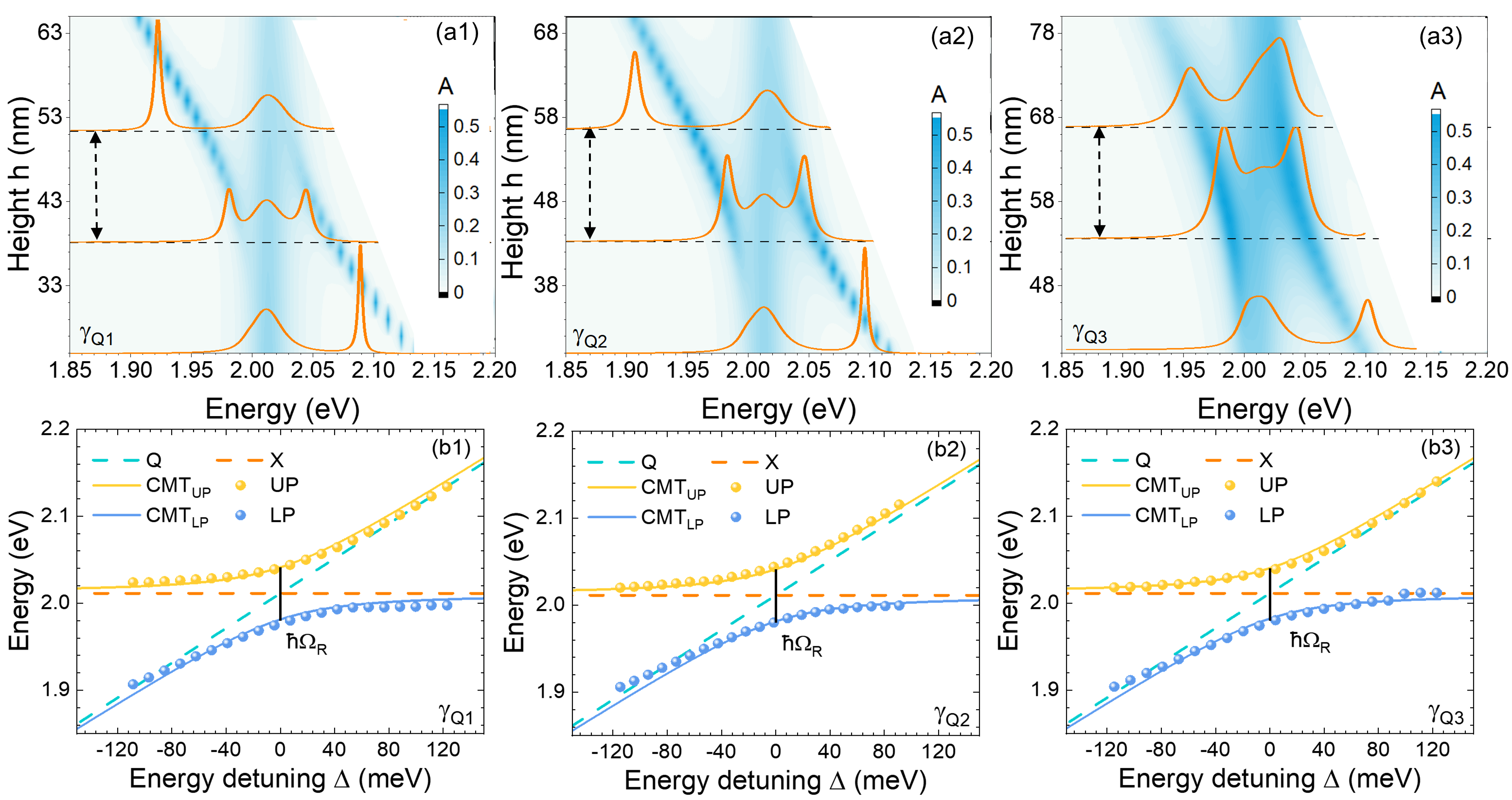}
\caption{\label{fig:pdis}(a) Polariton dispersion for (a1) $\rm{\hbar\gamma_{Q1}}= 1.5$ meV, $\rm y_{max}$ = 80 nm, (a2) $\rm{\hbar\gamma_{Q2}}=11.4$ meV, $\rm y_{max}$ = 100 nm, and (a3) $\rm{\hbar\gamma_{Q3}}=13.4$ meV, $\rm y_{max}$ = 144 nm, respectively. The absorbance spectra as a function of the height of the disk. The height of the disk (as labeled) is swept to achieve varied Q-BIC resonances Q and
corresponding energy detuning, $\mathrm{\Delta = E_{Q}- E_{X}}$. The radius of the disk, r = 240 nm and the gap bottom
position, $\rm y_{min}$ = -20 nm, are kept constant. 
The simulated absorbance spectra shown on each dispersion map correspond to the cases $\mathrm{E_{Q}<E_{X}}$, $\mathrm{E_{Q}=E_{X}}$ and $\mathrm{E_{Q}>E_{X}}$, respectively. The black dash line denotes the absorbance scale of 0.5. (b) Peak position as a function of energy detuning for (b1) $\rm \hbar\gamma_{Q1}= 1.5$ meV, (b2) $\rm \hbar\gamma_{Q2}=5.7$ meV and (b3) $\rm \hbar\gamma_{Q3}=13.4$ meV respectively. The dashed red line
represents the exciton energy (X), the dashed blue curve represents as the tuned Q-BIC cavity resonance (Q), and the two solid curves
(CMT$_{\mathrm{LP}}$ and CMT$_{\mathrm{UP}}$) are fitted polariton dispersion of lower polariton (LP) and upper polariton (UP) branches by using coupled mode theory (CMT) or Eq. 3. The normal mode splitting shown by the black solid line, is $\rm \hbar\Omega_R = 64.1$ meV, $\rm \hbar\Omega_R = 63.8$ meV, and $\rm \hbar\Omega_R = 58.9$ meV, respectively.}

\end{figure*}

\section{Theory}

As seen in Fig.~\ref{fig:sch} (a1), a hybrid cavity-emitter system is shown with the example of two-ports for incoming and outgoing electromagnetic waves. They system can be described by the temporal coupled-mode theory \cite{fan2003temporal}. The system is driven externally with a coupling constant $\rm d$. $\rm |Q|$, $\rm |X|$ denote the amplitude of photonic mode and excitonic mode respectively. The corresponding resonant frequency and damping rate are $\rm \omega_Q$ ($\rm \omega_X$) and $\rm \gamma_Q$ ($\rm \gamma_X$), respectively. The incoming wave amplitudes $\rm |S^{+}> = (S_1^+, S_2^+)$, and the corresponding outgoing wave amplitudes $\rm |S_1^->$, $\rm |S_2^->$, are then related by the following equations:

\begin{equation}
\label{matrix}
\begin{split}
     \rm \frac{dX}{dt}&=\rm (i\omega_X-\gamma_X)X+i \Omega Q, \\
     \rm \frac{dQ}{dt}&=\rm (i\omega_Q-\gamma_Q)Q+i \Omega X +d^T|s^+>, \\
    \rm |S^-> & =\rm C|S^+>+dQ.
\end{split}
\end{equation}
where the incident and outgoing waves are connected as $\rm |S^-> = S(\omega)|S^+>$. $\rm d$ is the coupling constant arranged in vectors. $\rm T$ denotes the transposing the row
vectors into column vectors. $\rm |Q|^2$ and $\rm |X|^2$ represent the corresponding stored electromagnetic energy. 
The integration yields:
\begin{equation}
\rm S(\omega)=C-\frac{i(\omega-\omega_Q)+\gamma_X}{(\omega-\omega_+)(\omega-\omega_-)}D
\end{equation}
where $\rm C$ is the background scattering matrix and $\rm D$ gives the coupling constant between resonance modes. The explicit expressions of $\rm C$ and $\rm D$ can be found in Ref. \cite{fan2003temporal}.

By calculating Det $\rm S(\omega)$ = 0, the upper and lower branches of energy are 

\begin{equation}
\label{Edispersion}
\begin{split}
   \rm  E_{\pm}&= \rm  \hbar{\omega}_{\pm}= \rm \frac{\hbar}{2}[{{\omega}_X+{\omega}_Q}+i{(\gamma_Q+\gamma_X)}]\\&
   \rm  \pm \frac{\hbar}{2} \sqrt{4\Omega^2+[(\omega_X-\omega_Q)+i(\gamma_X-\gamma_Q)]^2}
\end{split}
\end{equation}
The vacuum Rabi splitting, defined as the minimum energy spacing between the two branches, is $\rm \hbar\Omega_R = Re (E_{+}-E_{-})_{min}$. When $\rm  \omega_X=\omega_Q$ is met, it yields $\rm \rm \hbar\Omega_R=\hbar\sqrt{4\Omega^2-(\gamma_Q-\gamma_X)^2}$. To guarantee the coherent and reversible energy transfer, the energy anti-crossing behavior can only be resolved when the Rabi splitting is larger than the total dissipation energy of the hybrid system, $\rm \hbar\Omega_R> \hbar(\gamma_Q+\gamma_X)$ \cite{savona1995quantum,deng2010exciton,zhang2018photonic,peng2020separation}. Accordingly, the criteria of strong coupling or polaritonic coupling is

\begin{equation}
\label{criteria}
\rm \hbar\Omega>\frac{\hbar}{2}|\gamma_Q-\gamma_X|,
\rm \hbar\Omega>\hbar\sqrt{\frac{1}{2}(\gamma_Q^2+\gamma_X^2)}
\end{equation}
A special case exists when $\rm \omega_Q=\omega_X$ and $\rm \gamma_X=\gamma_Q$ are simultaneously met. Eq. ~\ref{Edispersion} yields, $\rm E_{\pm}=\hbar(\omega_X+ i\gamma_X\pm \Omega)$ or $\rm E_{\pm}=\hbar(\omega_Q+ i\gamma_Q \pm \Omega)$, implying the cavity/emitter's resonance splitting. The derived Rabi splitting energy becomes, $\rm \hbar\Omega_R = 2\hbar\Omega$. The criteria of strong coupling according to Eq.~\ref{criteria} becomes $\rm \hbar\Omega
>\hbar\gamma_Q=\hbar\gamma_X$.


Next we consider the critical coupling for the case of  a WS$_2$ monolayer with a Q-BIC optical resonator. In the case  with input only from a single port the energy stored in the optical resonator has a Lorentzian profile with the form $\rm |a|^2=\gamma_0/[(\omega-\omega_0)^2+\gamma_0^2]$, 
where $a$ is the amplitude, $\omega_0$ is the resonant frequency and $\gamma_0$ is the damping rate respectively. $\gamma_0$ has the contributions of radiative damping (scattering) and nonradiative damping (dephasing, dissipation or absorption) rates. The bare Q-BIC cavity has only the radiative scattering rate without dissipative loss. The monolayer WS$_2$ can be treated as only introducing additional dissipative loss to the Q-BIC cavity without breaking the mirror symmetry of the cavity resonances due to the ultra-thin dimension of monolayer \cite{piper2014total}. The absorbance of the symmetric hybrid system with one-port light incidence is $\rm A= [1-|detS(\omega)|^2]/2$ \cite{zanotto2014perfect,Zanotto2016}, 
which has a maximum absorbance value of 0.5 and sets the criteria for critical coupling. In this work, the fulfillment of the conditions for strong coupling on $\rm \hbar\Omega_R$ and $\rm \hbar\Omega$ while also achieving a maximum absorbance of 0.5 is referred to polaritonic critical coupling. The condition for maximum absorbance occurs when the radiative damping rate of the system matches the total of the non-radiative rates which can have a contribution due to non-radiative cavity losses as well as the monolayer WS$_2$ \cite{zanotto2014perfect}.

\begin{figure}[htbp]
\includegraphics[width=0.9\linewidth]{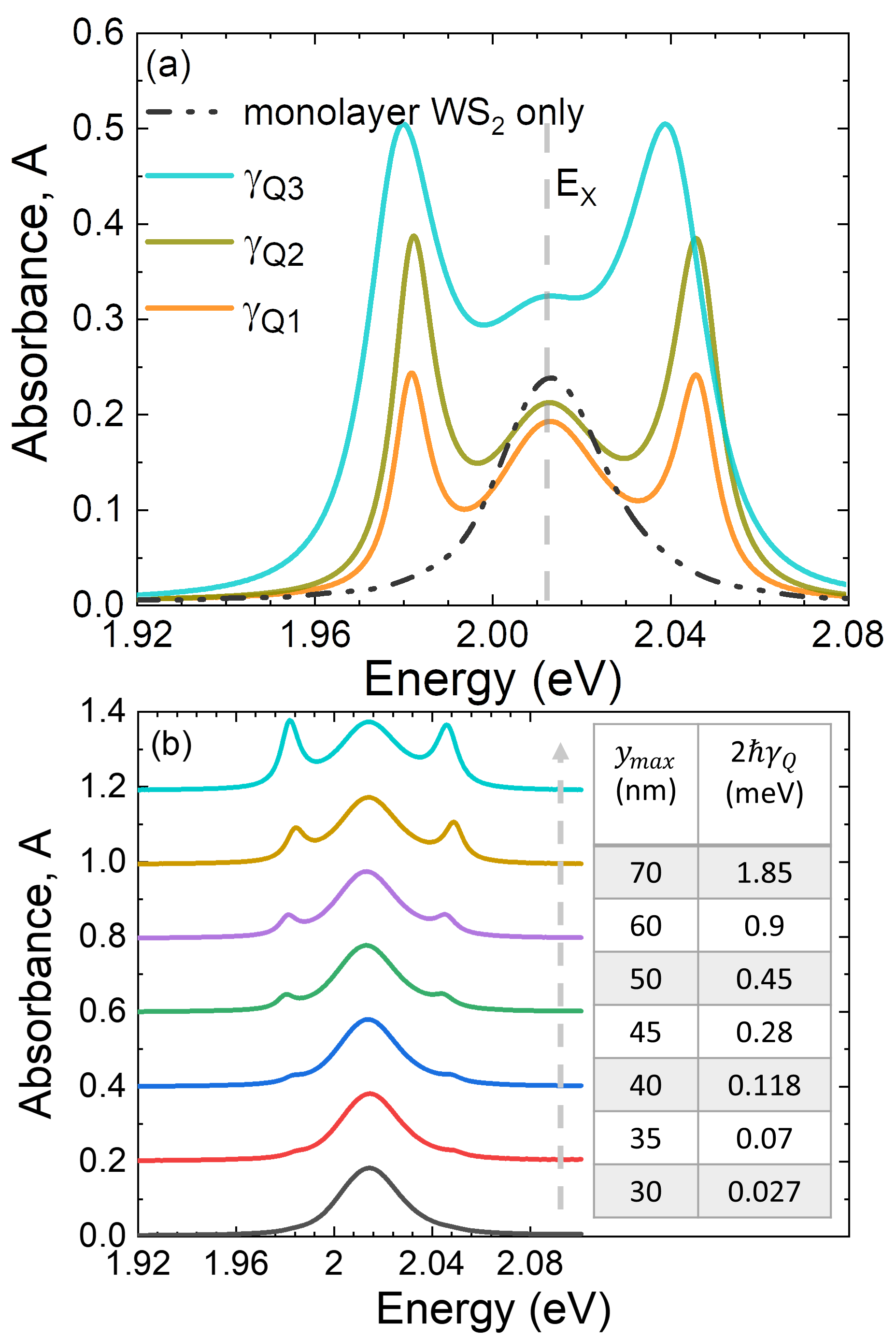}
\caption{\label{abs} (a) The calculated absorbance spectra of a bare WS$_2$ monolayer and the hybrid monolayer WS$_2$-Q-BIC cavities for $\rm{\hbar\gamma_{Q1}}= 1.5$ meV, $\rm{\hbar\gamma_{Q2}}=5.7$ meV and  $\rm{\hbar\gamma_{Q3}}=13.4$ meV respectively at the Rabi splitting energy. The grey dash line denotes the excitonic peak, $\rm E_x$. (b) The appearance of polaritonic splitting vs the damping rate of the Q-BIC cavities $2\rm{\hbar\gamma_{Q}}$: the stacked calculated absorbance of the hybrid monolayer WS$_2$-Q-BIC cavities with the value of $y_{max}$ and $2\rm{\hbar\gamma_{Q}}$ shown in inset. The grey dash line indicates the increasing $y_{max}$. }
\end{figure}

\begin{figure*}[htbp]
\includegraphics[width=1\linewidth]{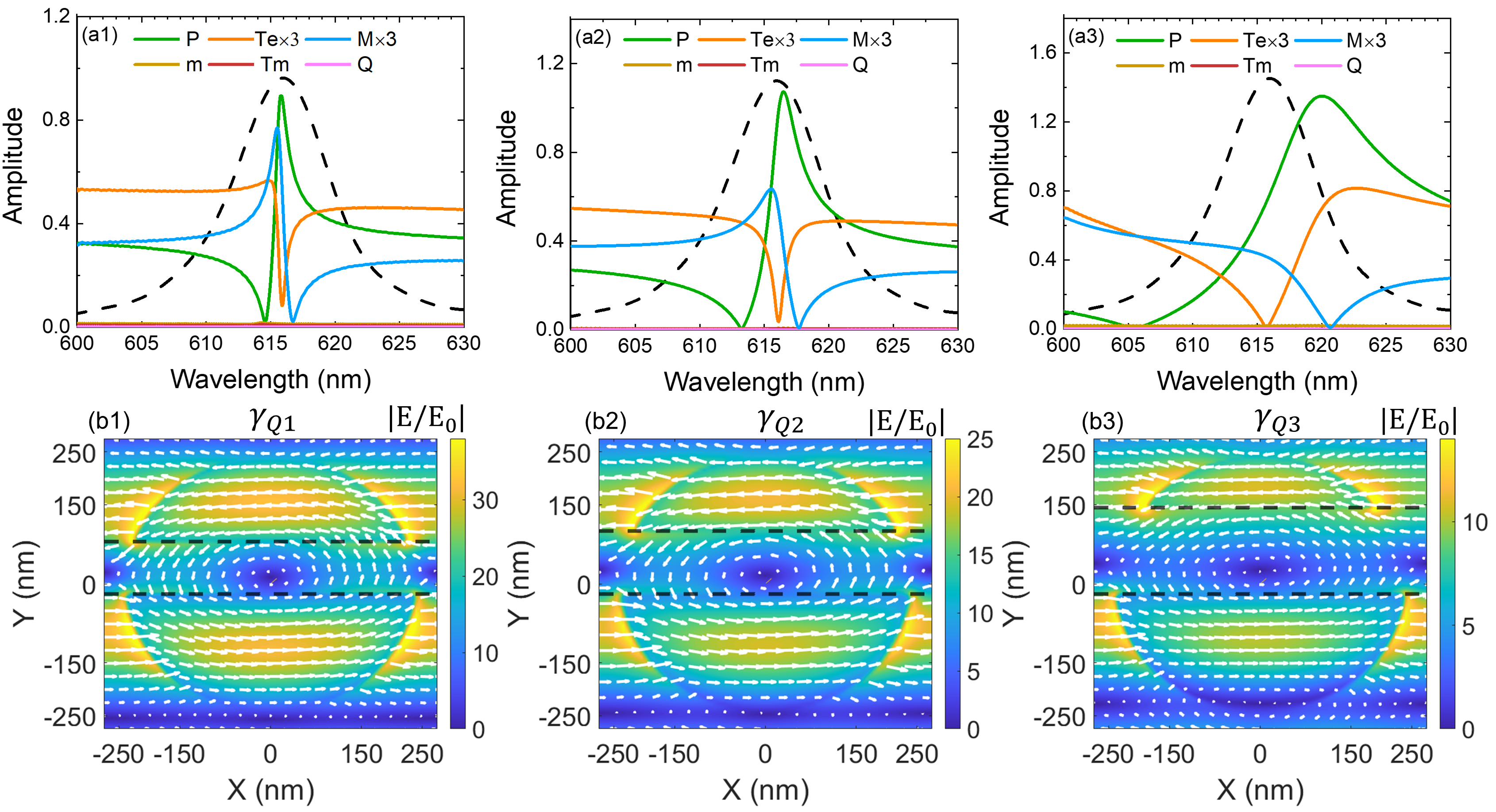}
\caption{\label{fig:field} The amplitude of the decomposed multipolar contributions contributing to the Q-BIC modes for (a1) $\rm{\hbar\gamma_{Q1}}= 1.5$ meV, (a2) $\rm{\hbar\gamma_{Q2}}=5.7$ meV and (a3) $\rm{\hbar\gamma_{Q3}}=13.4$ meV respectively, including electric dipole $\rm \textbf{P}$, electric toroidal dipole $\rm \textbf{T}e$, magnetic dipole $\rm \textbf{m}$, magnetic toroidal dipole $\rm \textbf{T}m$, electric quadrupole $\rm \textbf{Q}$ and magnetic quadrupole $\rm \textbf{M}$. The absorbance spectra (arbitrary units) of monolayer WS$_2$ is also shown as black dash line, below which illustrates the relative spectral linewidth of the exciton and Q-BIC cavity's contributing modes. (b1-b3) The corresponding relative amplitude of the electric field, $\rm|\textbf{E}/\textbf{E}_0|$ in the $x-y$ plane through the middle of the Q-BIC cavities. The arrows denote the electric field vectors. The inspected wavelength is at the excitonic peak, $\rm E_x$ = 2.012 eV or 616 nm. The black dash line indicates the position of the air gap, $\rm y_{min}$ is kept as -20 nm,  $\rm y_{max}$ = 80 nm for $\rm \hbar\gamma_{Q1}$, $\rm y_{max}$ = 100 nm for $\rm \hbar\gamma_{Q2}$ and $\rm y_{max}$ = 144 nm for $\rm \hbar\gamma_{Q3}$.}
\end{figure*}


\section{Polaritonic Critical Coupling Realization}

Fig.~\ref{fig:sch} (a2) shows a schematic of the hybrid structure, where the two-level monolayer WS$_2$ is in contact with the Q-BIC cavity. The complex dielectric permittivity of the monolayer WS$_2$, as a function of the photon energy $\rm E$ is 
\begin{equation}
      \rm \epsilon(E)=\epsilon_B+\sum_{j=1}^{n}\frac{f_j}{E_{0j}^2-E^2-i\Gamma_{j}E}
\end{equation}
where $\rm \epsilon_B$ denotes the dielectric permittivity of the background. $\rm E_{0j}$, $\rm f_j$, and $\rm \Gamma_{j}$ are the resonance energy, oscillator strengths, and the damping rate of the oscillator with index $\rm j$, respectively. The fit parameters as well as monolayer thickness are taken from Ref. \cite{li2014measurement, Goncalves2018} with details listed in Table I.

\begin{table}[h!]
\begin{tabular}{ |p{1cm}||p{2cm}|p{2cm}|p{2cm}| }
\hline
 \multicolumn{4}{|c|}{ $\rm \epsilon_B$=8.76} \\
 \hline
 j & $\rm f_j([eV]^2)$ &  $\rm E_{0j} (eV)$& $\rm\Gamma_j (eV)$ \\
 \hline

 1.   & 1.9    &2.014   &   0.029\\
 2.   & 0.254  & 2.185   & 0.1\\
 3.   &0.146  & 2.25     &  0.1\\
 4.    &0.068   & 2.285     & 0.1\\
 5.   &   3.07  & 2.402      &0.14\\
6.     & 1.17   & 2.575      &0.21\\
 7.   & 0.068  & 2.655       &0.21\\
8.    & 15.5    & 2.845    &0.265\\
 9.    & 12.7  & 3.047    &0.25\\
 \hline

\end{tabular}
\caption{The fitting parameters for the permittivity of monolayer WS$_2$.}
\label{table1}
\end{table}

As seen in Fig.~\ref{fig:sch} (b1), the damping rate of A-exciton resonance is determined from the calculated absorbance spectrum, $\rm A = 1 - R - T$, where a Lorentzian fit is performed. $\rm R$, $\rm T$ are obtained by Lumerical finite-difference time-domain (FDTD) simulation, where periodic boundary conditions are applied over $x-y$ directions and perfectly matched layer boundary condition is applied over $z$ direction. The extracted spectral linewidth, is $2\hbar\gamma_X$ = 26.8 meV. Correspondingly, $\rm \hbar\gamma_X$ = 13.4 meV, results from the homogeneous broadening driven by the radiative and nonradiative decay rates, the inhomogeneous broadening is negligible \cite{wang2016coherent}.

As seen in Fig.~\ref{fig:sch} (a3), the designed cavity is a Q-BIC gap cavity metasurface, where an asymmetric air gap is inserted in the TiO$_2$ disk resonators. The explored geometry including the periodicity, the gap position, the disk radius as well as the height have been selected to guarantee tuning on and off the spectral overlap of the resonance mode with the excitonic resonance to enable energy detuning. The permittivity of TiO$_2$ is taken from the experimental data in Ref. \cite{sarkar2019hybridized}. More details of the designed gap cavity as well as its Q-BIC nature can be seen in Ref. \cite{zhang2021ultra}. As the WS$_2$ monolayer excitonic resonance energy, $E_X$ is constant, the cavity resonance, $\rm E_{Q}$ is varied by sweeping only the height of the designed Q-BIC resonators to enable energy detuning. It is necessary to note that sweeping only the height while maintaining the asymmetry parameter fixed does not affect the damping rate. Generally, the absorbed power of a structure, here the cavity or the WS$_2$ monolayer, can be quantified by $
    \rm P_{abs}=\frac{1}{2}\iiint |\bm{\textbf{E}}|^2Im(\epsilon) dV$ \cite{jackson1999classical},
where $\rm |\bm{\textbf{E}}|$ is the amplitude of the electric field within the cavity or monolayer WS$_2$, $\rm \epsilon$ is the corresponding permittivity. The absorbance is in principle the ratio of the
total absorbed power within a volume $\rm V$ to the incoming power through the exposed surface area \cite{jackson1999classical,Baffou2009}. Since the Q-BIC gap cavity has an infinitesimal imaginary part of the permittivity, the absorbance or dissipation rate of the Q-BIC cavity is negligible. Therefore the damping rate, $\gamma_Q$, of the Q-BIC cavity purely results from the radiative scattering. The calculated absorbance in Fig.~\ref{fig:sch} (b2) also confirms negligible dissipation losses in the Q-BIC cavity. The absorbance of the hybrid structure results only from the dissipation losses due to the monolayer WS$_2$, which is affected by the coupling with the cavity's resonant modes. To explore the coupling between the Q-BIC cavity and monolayer WS$_2$, the damping rate of the Q-BIC cavity is extracted from a Fano fit of the transmission spectrum, an example can be seen in Fig.~\ref{fig:sch} (b2). The Fano fit is performed by $\rm T(E)=T_0+A_0\frac{[q+2(E-E_0)/E_w]^2}{1+[2(E-E_0)/E_w]^2}$, where $\rm E_0$ is the resonant energy, $\rm E_w$ is the energy linewidth, or full-width at half-maximum. $\rm T_0$ is the transmission offset, $\rm A_0$ is the continuum-discrete coupling parameter and $\rm q$ is the Breit-Wigner-Fano parameter. 

The calculated absorbance for three damping rates of the Q-BIC cavity are shown in Fig.~\ref{fig:pdis} (a1-a3) respectively. The damping rate is determined by the asymmetry parameter, which standing in air is shown in Fig.~\ref{fig:pdis} (b3). Tuning of the cavity resonance with respect to the exciton energy can be achieved by varying the disk height. As mentioned earlier varying $h$ does not affect the damping rate. Fig.~\ref{fig:pdis} (a1-a3) show colour maps of the absorbance as a function of energy with varying disk height. Three spectra are shown in each case corresponding to the cases $\rm E_Q < E_X$,  $\rm E_Q = E_X$, and $\rm E_Q > E_X$. Due to its high refractive index, the monolayer $\rm WS_2$ causes an energy red-shifting of the Q-BIC mode \cite{Karanikolas2020}. To take account of this effect, the energy of the Q-BIC mode is simulated by replacing the monolayer WS$_2$ with an ultra-thin nanosheet, which has the same thickness as monolayer $\rm WS_2$ but with an average refractive index, $\rm n = 4.5$ estimated from Ref. \cite{li2014measurement,cao2020normal}, which also agrees with the experimental determination in Ref. \cite{Hsu2019}. The energy of Q-BIC mode of the metasurface hybridized with the layer with $\rm n = 4.5$ ($\rm E_Q$) as well as excitonic energy ($\rm E_X$) can be seen in Fig.~\ref{fig:pdis} (b1-b3). The anti-crossing behavior is seen in (b1) for $\rm \hbar\gamma_{Q1}= 1.5$ meV, (b2) for $\rm \hbar\gamma_{Q2}=5.7$ meV and (b3) for $\rm \hbar\gamma_{Q3}=$ 13.4 meV, which is a signature of polaritonic states generation. The upper (UP) and lower (LP) polariton energy branches are well reproduced by Eq.~\ref{Edispersion}.  The Rabi splitting energy, corresponding to the minimum value of energy separation between the two branches, indicated by the black line, are (b1) $\rm \hbar\Omega_R = 64.1$ meV, (b2) $\rm \hbar\Omega_R = 63.8$ meV, (b3) $\rm \hbar\Omega_R = 58$ meV. Comparing the values of $\rm \hbar\Omega_R$, $\rm \hbar\gamma_{Q}$ and $\rm \hbar\gamma_{X}$, it yields, $\rm \hbar\Omega_R>\hbar(\gamma_Q+\gamma_X)$. Furthermore, the extracted coupling constant according to Eq.~\ref{Edispersion} is (b1) $\hbar\Omega_1 = 32.5$ meV, (b2) $\rm \hbar\Omega_2 = 32.1$ meV, (b3) $\rm \hbar\Omega_3 = 29.5$ meV, which all clearly meet the criteria, Eq.~\ref{criteria}: $\rm \hbar\Omega>\hbar|\gamma_Q-\gamma_X|/{2}$ and $\rm \hbar\Omega >\hbar\sqrt{\frac{1}{2}(\gamma_Q^2+\gamma_X^2)}$. Therefore, polaritonic coupling has been achieved in the hybrid system.

Note that the excitonic absorbance peak, $\rm E_X$ = 2.012 eV or 616 nm remains visible in the spectra in Fig.~\ref{fig:pdis} (a1-a3) due to the presence of non-hybridized excitons remaining in the system \cite{piper2014total, Kang2018}.
For a closer inspection of the absorbance value, the absorbance spectra of the bare and the hybrid structure comprised of monolayer WS$_2$ with the Q-BIC cavities for the three damping rates of the Q-BIC cavity can be seen in Fig.~\ref{abs} (a). It is clear that non-hybridized excitons are always present and only some of the exciton population form polaritons, in particular for the damping rates of $\rm \gamma_{Q1}$ and $\rm \gamma_{Q2}$. Furthermore, for the case $\rm \gamma_{Q3}=\gamma_{X}$, the absorbance value increases dramatically compared to the bare WS$_2$ monolayer, which implies that the population of the excitonic states or the absorbance at $\rm E_X$ is enhanced due to the resonance of the cavity. Particularly, as can be seen in Fig.~\ref{fig:pdis} (a3), the absorbance value at the two 
polariton energies reaches 0.5, which is the maximum possible value that can be obtained for a single input beam, occurring when the condition for strong critical coupling or polaritonic critical coupling is satisfied, namely when the radiation rate matches the dissipative non-radiative rate in the coupled system. This observation indicates that it should be possible to achieve coherent perfect absorption using two input ports \cite{zanotto2014perfect}. It is interesting to further inspect the effect of the Q-BIC cavity's linewidth on the polaritonic coupling, with results shown in Fig.~\ref{abs} (b). It is clear that with increasing damping rates by tuning $\rm y_{max}$, as indicated by the grey dash line, the signature of the polaritonic coupling appears and gradually displays a pronounced Rabi splitting. It shows that a minimum required energy linewidth of the cavity for polaritonic coupling with the excitons, is at around $\rm y_{max} = 40 nm$ and $2\hbar\gamma_X$ = 0.118 meV.

\section{Multipolar Decomposition of Q-BIC Modes}

To get a deeper physical insight into the multipolar modes driving the absorbance enhancement and polaritonic critical coupling, multipolar decomposition is performed following Ref. \cite{evlyukhin2016optical}, where the electric field is integrated over one unit cell of an array with employed periodic boundary conditions along the $x-y$ directions and over
the height of the Q-BIC cavity along the $z$ direction. The amplitude of decomposed multipolar modes contributing to the reflectance/transmittance coefficient of the array are shown in Fig.~\ref{fig:field} (a1-a3) for $\rm \hbar\gamma_{Q1}= 1.5$ meV, $\rm \hbar\gamma_{Q2}=5.7$ meV and $\rm \hbar\gamma_{Q3}=$ 13.4 meV,  respectively. The excitonic absorbance spectrum of monolayer WS$_2$ is also shown to illustrate the spectral width or relative damping rates of the monolayer WS$_2$ and Q-BIC modes. It is clear that the total electric dipole, which include the contributions of electric dipole, $\rm \textbf{P}$ and electric toroidal dipole, $\rm \textbf{T}e$, together with the contribution from magnetic quadrupole $\rm \textbf{M}$ dominate the radiative damping rate of the cavity. The magnetic dipole $\textbf{m}$, magnetic toroidal dipole $\rm \textbf{T}m$, and electric quadrupole $\textbf{Q}$ are negligible. Moreover, the gradual broadening of the dominant multipolar modes explains the increasing damping rate of Q-BIC modes \cite{koshelev2018asymmetric}. The electric field distribution at the inspected wavelength, which corresponds to the excitonic absorbance peak at $\rm E_x$ = 2.012 eV or 616 nm, as well as the field vectors can be seen in Fig.~\ref{fig:field} (b1-b3). With increasing gap width, or $\rm y_{max}$, the electric field amplitude decreases, and the trend agrees with the reducing Rabi splitting energy with increasing radiative damping rate. Since the dominant electric dipole, $\rm \textbf{P}$, electric toroidal dipole, $\rm \textbf{T}e$, as well as magnetic quadrupole $\rm \textbf{M}$ have an even parity in the forward and backward scattering plane \cite{evlyukhin2016optical,PhysRevB.103.195419}, the proposed Q-BIC gap resonator has a geometric mirror symmetry as well as radially scattering symmetry along $z$ direction. The damping rate of the Q-BIC mode is purely driven by the in-plane geometric symmetry breaking along the $y$ direction.

\section{Summary}

Inspired by a fundamental question within the framework of light-matter coupling regarding maximizing the absorbance of electromagnetic energy for a polaritonic system, a metasurface Q-BIC cavity is proposed for polaritonic critical coupling, where critical coupling and strong coupling are simultaneously realized. By considering a system comprised of a TiO$_2$ Q-BIC metasurface optical cavity resonator coupled with the excitons of monolayer WS$_2$ under single-beam excitation, we have explored the conditions for achieving polaritonic critical coupling. The Q-BIC cavity enables the manipulation of the damping rates, where the radiative scattering dominates. Through manipulating the relative damping rate of the Q-BIC cavity and WS$_2$ monolayer, it is demonstrated that strong coupling and polaritonic critical coupling with absorbance enhancement can be realized for different radiative damping rates. The maximum possible absorbance of 0.5 in the strong coupling regime is observed when the conditions for  polaritonic critical coupling are met. The underlying driving multipolar modes are also explored, revealing that the total electric dipole and magnetic quadrupole dominate the cavity's radially-symmetric damping rate. Our study on a WS$_2$ monolayer interacting with a Q-BIC resonator with a view to polariton physics may deepen more understanding of absorbance 
manipulation in the strong coupling regime, spur studies of coherent perfect absorption in a polariton system, and further guide the realization of the efficient polaritonic devices.

\begin{acknowledgments}
We wish to acknowledge the support of Science Foundation Ireland (SFI) under Grant Number 16/IA/4550. 
\end{acknowledgments}

\appendix

\nocite{*}

\bibliography{apssamp}

\begin{thebibliography}{64}%
\makeatletter
\providecommand \@ifxundefined [1]{%
 \@ifx{#1\undefined}
}%
\providecommand \@ifnum [1]{%
 \ifnum #1\expandafter \@firstoftwo
 \else \expandafter \@secondoftwo
 \fi
}%
\providecommand \@ifx [1]{%
 \ifx #1\expandafter \@firstoftwo
 \else \expandafter \@secondoftwo
 \fi
}%
\providecommand \natexlab [1]{#1}%
\providecommand \enquote  [1]{``#1''}%
\providecommand \bibnamefont  [1]{#1}%
\providecommand \bibfnamefont [1]{#1}%
\providecommand \citenamefont [1]{#1}%
\providecommand \href@noop [0]{\@secondoftwo}%
\providecommand \href [0]{\begingroup \@sanitize@url \@href}%
\providecommand \@href[1]{\@@startlink{#1}\@@href}%
\providecommand \@@href[1]{\endgroup#1\@@endlink}%
\providecommand \@sanitize@url [0]{\catcode `\\12\catcode `\$12\catcode
  `\&12\catcode `\#12\catcode `\^12\catcode `\_12\catcode `\%12\relax}%
\providecommand \@@startlink[1]{}%
\providecommand \@@endlink[0]{}%
\providecommand \url  [0]{\begingroup\@sanitize@url \@url }%
\providecommand \@url [1]{\endgroup\@href {#1}{\urlprefix }}%
\providecommand \urlprefix  [0]{URL }%
\providecommand \Eprint [0]{\href }%
\providecommand \doibase [0]{https://doi.org/}%
\providecommand \selectlanguage [0]{\@gobble}%
\providecommand \bibinfo  [0]{\@secondoftwo}%
\providecommand \bibfield  [0]{\@secondoftwo}%
\providecommand \translation [1]{[#1]}%
\providecommand \BibitemOpen [0]{}%
\providecommand \bibitemStop [0]{}%
\providecommand \bibitemNoStop [0]{.\EOS\space}%
\providecommand \EOS [0]{\spacefactor3000\relax}%
\providecommand \BibitemShut  [1]{\csname bibitem#1\endcsname}%
\let\auto@bib@innerbib\@empty
\bibitem [{\citenamefont {K{\"u}hn}\ \emph {et~al.}(2006)\citenamefont
  {K{\"u}hn}, \citenamefont {H{\aa}kanson}, \citenamefont {Rogobete},\ and\
  \citenamefont {Sandoghdar}}]{PhysRevLett.97.017402}%
  \BibitemOpen
  \bibfield  {author} {\bibinfo {author} {\bibfnamefont {S.}~\bibnamefont
  {K{\"u}hn}}, \bibinfo {author} {\bibfnamefont {U.}~\bibnamefont
  {H{\aa}kanson}}, \bibinfo {author} {\bibfnamefont {L.}~\bibnamefont
  {Rogobete}},\ and\ \bibinfo {author} {\bibfnamefont {V.}~\bibnamefont
  {Sandoghdar}},\ }\href {https://doi.org/10.1103/PhysRevLett.97.017402}
  {\bibfield  {journal} {\bibinfo  {journal} {Phys. Rev. Lett.}\ }\textbf
  {\bibinfo {volume} {97}},\ \bibinfo {pages} {017402} (\bibinfo {year}
  {2006})}\BibitemShut {NoStop}%
\bibitem [{\citenamefont {Novotny}\ and\ \citenamefont
  {Van~Hulst}(2011)}]{novotny2011antennas}%
  \BibitemOpen
  \bibfield  {author} {\bibinfo {author} {\bibfnamefont {L.}~\bibnamefont
  {Novotny}}\ and\ \bibinfo {author} {\bibfnamefont {N.}~\bibnamefont
  {Van~Hulst}},\ }\href
  {https://doi.org/https://doi.org/10.1038/nphoton.2010.237} {\bibfield
  {journal} {\bibinfo  {journal} {Nat. Photonics}\ }\textbf {\bibinfo {volume}
  {5}},\ \bibinfo {pages} {83} (\bibinfo {year} {2011})}\BibitemShut {NoStop}%
\bibitem [{\citenamefont {T{\"o}rm{\"a}}\ and\ \citenamefont
  {Barnes}(2014)}]{torma2014strong}%
  \BibitemOpen
  \bibfield  {author} {\bibinfo {author} {\bibfnamefont {P.}~\bibnamefont
  {T{\"o}rm{\"a}}}\ and\ \bibinfo {author} {\bibfnamefont {W.~L.}\ \bibnamefont
  {Barnes}},\ }\href {https://doi.org/10.1088/0034-4885/78/1/013901} {\bibfield
   {journal} {\bibinfo  {journal} {Rep. Prog. Phys.}\ }\textbf {\bibinfo
  {volume} {78}},\ \bibinfo {pages} {013901} (\bibinfo {year}
  {2014})}\BibitemShut {NoStop}%
\bibitem [{\citenamefont {Curto}\ \emph {et~al.}(2010)\citenamefont {Curto},
  \citenamefont {Volpe}, \citenamefont {Taminiau}, \citenamefont {Kreuzer},
  \citenamefont {Quidant},\ and\ \citenamefont {van
  Hulst}}]{curto2010unidirectional}%
  \BibitemOpen
  \bibfield  {author} {\bibinfo {author} {\bibfnamefont {A.~G.}\ \bibnamefont
  {Curto}}, \bibinfo {author} {\bibfnamefont {G.}~\bibnamefont {Volpe}},
  \bibinfo {author} {\bibfnamefont {T.~H.}\ \bibnamefont {Taminiau}}, \bibinfo
  {author} {\bibfnamefont {M.~P.}\ \bibnamefont {Kreuzer}}, \bibinfo {author}
  {\bibfnamefont {R.}~\bibnamefont {Quidant}},\ and\ \bibinfo {author}
  {\bibfnamefont {N.~F.}\ \bibnamefont {van Hulst}},\ }\href
  {https://doi.org/10.1126/science.1191922} {\bibfield  {journal} {\bibinfo
  {journal} {Science}\ }\textbf {\bibinfo {volume} {329}},\ \bibinfo {pages}
  {930} (\bibinfo {year} {2010})}\BibitemShut {NoStop}%
\bibitem [{\citenamefont {Tr\"ugler}\ and\ \citenamefont
  {Hohenester}(2008)}]{PhysRevB.77.115403}%
  \BibitemOpen
  \bibfield  {author} {\bibinfo {author} {\bibfnamefont {A.}~\bibnamefont
  {Tr\"ugler}}\ and\ \bibinfo {author} {\bibfnamefont {U.}~\bibnamefont
  {Hohenester}},\ }\href
  {https://doi.org/https://doi.org/10.1103/PhysRevB.77.115403} {\bibfield
  {journal} {\bibinfo  {journal} {Phys. Rev. B}\ }\textbf {\bibinfo {volume}
  {77}},\ \bibinfo {pages} {115403} (\bibinfo {year} {2008})}\BibitemShut
  {NoStop}%
\bibitem [{\citenamefont {Bellessa}\ \emph
  {et~al.}(2004{\natexlab{a}})\citenamefont {Bellessa}, \citenamefont
  {Bonnand}, \citenamefont {Plenet},\ and\ \citenamefont
  {Mugnier}}]{PhysRevLett.93.036404}%
  \BibitemOpen
  \bibfield  {author} {\bibinfo {author} {\bibfnamefont {J.}~\bibnamefont
  {Bellessa}}, \bibinfo {author} {\bibfnamefont {C.}~\bibnamefont {Bonnand}},
  \bibinfo {author} {\bibfnamefont {J.}~\bibnamefont {Plenet}},\ and\ \bibinfo
  {author} {\bibfnamefont {J.}~\bibnamefont {Mugnier}},\ }\href
  {https://doi.org/https://doi.org/10.1103/PhysRevLett.93.036404} {\bibfield
  {journal} {\bibinfo  {journal} {Phys. Rev. Lett.}\ }\textbf {\bibinfo
  {volume} {93}},\ \bibinfo {pages} {036404} (\bibinfo {year}
  {2004}{\natexlab{a}})}\BibitemShut {NoStop}%
\bibitem [{\citenamefont {Ren}\ \emph {et~al.}(2017)\citenamefont {Ren},
  \citenamefont {Gu}, \citenamefont {Zhao}, \citenamefont {Zhang},
  \citenamefont {Zhang},\ and\ \citenamefont {Gong}}]{PhysRevLett.118.073604}%
  \BibitemOpen
  \bibfield  {author} {\bibinfo {author} {\bibfnamefont {J.}~\bibnamefont
  {Ren}}, \bibinfo {author} {\bibfnamefont {Y.}~\bibnamefont {Gu}}, \bibinfo
  {author} {\bibfnamefont {D.}~\bibnamefont {Zhao}}, \bibinfo {author}
  {\bibfnamefont {F.}~\bibnamefont {Zhang}}, \bibinfo {author} {\bibfnamefont
  {T.}~\bibnamefont {Zhang}},\ and\ \bibinfo {author} {\bibfnamefont
  {Q.}~\bibnamefont {Gong}},\ }\href
  {https://doi.org/https://doi.org/10.1103/PhysRevLett.118.073604} {\bibfield
  {journal} {\bibinfo  {journal} {Phys. Rev. Lett.}\ }\textbf {\bibinfo
  {volume} {118}},\ \bibinfo {pages} {073604} (\bibinfo {year}
  {2017})}\BibitemShut {NoStop}%
\bibitem [{\citenamefont {Chikkaraddy}\ \emph {et~al.}(2016)\citenamefont
  {Chikkaraddy}, \citenamefont {De~Nijs}, \citenamefont {Benz}, \citenamefont
  {Barrow}, \citenamefont {Scherman}, \citenamefont {Rosta}, \citenamefont
  {Demetriadou}, \citenamefont {Fox}, \citenamefont {Hess},\ and\ \citenamefont
  {Baumberg}}]{chikkaraddy2016single}%
  \BibitemOpen
  \bibfield  {author} {\bibinfo {author} {\bibfnamefont {R.}~\bibnamefont
  {Chikkaraddy}}, \bibinfo {author} {\bibfnamefont {B.}~\bibnamefont
  {De~Nijs}}, \bibinfo {author} {\bibfnamefont {F.}~\bibnamefont {Benz}},
  \bibinfo {author} {\bibfnamefont {S.~J.}\ \bibnamefont {Barrow}}, \bibinfo
  {author} {\bibfnamefont {O.~A.}\ \bibnamefont {Scherman}}, \bibinfo {author}
  {\bibfnamefont {E.}~\bibnamefont {Rosta}}, \bibinfo {author} {\bibfnamefont
  {A.}~\bibnamefont {Demetriadou}}, \bibinfo {author} {\bibfnamefont
  {P.}~\bibnamefont {Fox}}, \bibinfo {author} {\bibfnamefont {O.}~\bibnamefont
  {Hess}},\ and\ \bibinfo {author} {\bibfnamefont {J.~J.}\ \bibnamefont
  {Baumberg}},\ }\href {https://doi.org/https://doi.org/10.1038/nature17974}
  {\bibfield  {journal} {\bibinfo  {journal} {Nature}\ }\textbf {\bibinfo
  {volume} {535}},\ \bibinfo {pages} {127} (\bibinfo {year}
  {2016})}\BibitemShut {NoStop}%
\bibitem [{\citenamefont {Peter}\ \emph {et~al.}(2005)\citenamefont {Peter},
  \citenamefont {Senellart}, \citenamefont {Martrou}, \citenamefont
  {Lema{\^\i}tre}, \citenamefont {Hours}, \citenamefont {G{\'e}rard},\ and\
  \citenamefont {Bloch}}]{peter2005exciton}%
  \BibitemOpen
  \bibfield  {author} {\bibinfo {author} {\bibfnamefont {E.}~\bibnamefont
  {Peter}}, \bibinfo {author} {\bibfnamefont {P.}~\bibnamefont {Senellart}},
  \bibinfo {author} {\bibfnamefont {D.}~\bibnamefont {Martrou}}, \bibinfo
  {author} {\bibfnamefont {A.}~\bibnamefont {Lema{\^\i}tre}}, \bibinfo {author}
  {\bibfnamefont {J.}~\bibnamefont {Hours}}, \bibinfo {author} {\bibfnamefont
  {J.}~\bibnamefont {G{\'e}rard}},\ and\ \bibinfo {author} {\bibfnamefont
  {J.}~\bibnamefont {Bloch}},\ }\href
  {https://doi.org/10.1103/PhysRevLett.95.067401} {\bibfield  {journal}
  {\bibinfo  {journal} {Phys. Rev. Lett.}\ }\textbf {\bibinfo {volume} {95}},\
  \bibinfo {pages} {067401} (\bibinfo {year} {2005})}\BibitemShut {NoStop}%
\bibitem [{\citenamefont {Wang}\ \emph
  {et~al.}(2016{\natexlab{a}})\citenamefont {Wang}, \citenamefont {Nie},
  \citenamefont {Qin}, \citenamefont {Hu},\ and\ \citenamefont
  {Tang}}]{wang2016giant}%
  \BibitemOpen
  \bibfield  {author} {\bibinfo {author} {\bibfnamefont {Z.}~\bibnamefont
  {Wang}}, \bibinfo {author} {\bibfnamefont {J.}~\bibnamefont {Nie}}, \bibinfo
  {author} {\bibfnamefont {W.}~\bibnamefont {Qin}}, \bibinfo {author}
  {\bibfnamefont {Q.}~\bibnamefont {Hu}},\ and\ \bibinfo {author}
  {\bibfnamefont {B.~Z.}\ \bibnamefont {Tang}},\ }\href
  {https://doi.org/https://doi.org/10.1038/ncomms11283} {\bibfield  {journal}
  {\bibinfo  {journal} {Nat. Commun.}\ }\textbf {\bibinfo {volume} {7}},\
  \bibinfo {pages} {1} (\bibinfo {year} {2016}{\natexlab{a}})}\BibitemShut
  {NoStop}%
\bibitem [{\citenamefont {Michaeli}\ \emph {et~al.}(2017)\citenamefont
  {Michaeli}, \citenamefont {Keren-Zur}, \citenamefont {Avayu}, \citenamefont
  {Suchowski},\ and\ \citenamefont {Ellenbogen}}]{michaeli2017nonlinear}%
  \BibitemOpen
  \bibfield  {author} {\bibinfo {author} {\bibfnamefont {L.}~\bibnamefont
  {Michaeli}}, \bibinfo {author} {\bibfnamefont {S.}~\bibnamefont {Keren-Zur}},
  \bibinfo {author} {\bibfnamefont {O.}~\bibnamefont {Avayu}}, \bibinfo
  {author} {\bibfnamefont {H.}~\bibnamefont {Suchowski}},\ and\ \bibinfo
  {author} {\bibfnamefont {T.}~\bibnamefont {Ellenbogen}},\ }\href
  {https://doi.org/10.1103/PhysRevLett.118.243904} {\bibfield  {journal}
  {\bibinfo  {journal} {Phys. Rev. Lett.}\ }\textbf {\bibinfo {volume} {118}},\
  \bibinfo {pages} {243904} (\bibinfo {year} {2017})}\BibitemShut {NoStop}%
\bibitem [{\citenamefont {Koshelev}\ \emph {et~al.}(2020)\citenamefont
  {Koshelev}, \citenamefont {Kruk}, \citenamefont {Melik-Gaykazyan},
  \citenamefont {Choi}, \citenamefont {Bogdanov}, \citenamefont {Park},\ and\
  \citenamefont {Kivshar}}]{koshelev2020subwavelength}%
  \BibitemOpen
  \bibfield  {author} {\bibinfo {author} {\bibfnamefont {K.}~\bibnamefont
  {Koshelev}}, \bibinfo {author} {\bibfnamefont {S.}~\bibnamefont {Kruk}},
  \bibinfo {author} {\bibfnamefont {E.}~\bibnamefont {Melik-Gaykazyan}},
  \bibinfo {author} {\bibfnamefont {J.-H.}\ \bibnamefont {Choi}}, \bibinfo
  {author} {\bibfnamefont {A.}~\bibnamefont {Bogdanov}}, \bibinfo {author}
  {\bibfnamefont {H.-G.}\ \bibnamefont {Park}},\ and\ \bibinfo {author}
  {\bibfnamefont {Y.}~\bibnamefont {Kivshar}},\ }\href
  {https://doi.org/10.1126/science.aaz3985} {\bibfield  {journal} {\bibinfo
  {journal} {Science}\ }\textbf {\bibinfo {volume} {367}},\ \bibinfo {pages}
  {288} (\bibinfo {year} {2020})}\BibitemShut {NoStop}%
\bibitem [{\citenamefont {Bhattacharya}\ \emph {et~al.}(2014)\citenamefont
  {Bhattacharya}, \citenamefont {Frost}, \citenamefont {Deshpande},
  \citenamefont {Baten}, \citenamefont {Hazari},\ and\ \citenamefont
  {Das}}]{Bhattacharya2014}%
  \BibitemOpen
  \bibfield  {author} {\bibinfo {author} {\bibfnamefont {P.}~\bibnamefont
  {Bhattacharya}}, \bibinfo {author} {\bibfnamefont {T.}~\bibnamefont {Frost}},
  \bibinfo {author} {\bibfnamefont {S.}~\bibnamefont {Deshpande}}, \bibinfo
  {author} {\bibfnamefont {M.~Z.}\ \bibnamefont {Baten}}, \bibinfo {author}
  {\bibfnamefont {A.}~\bibnamefont {Hazari}},\ and\ \bibinfo {author}
  {\bibfnamefont {A.}~\bibnamefont {Das}},\ }\href
  {https://doi.org/https://doi.org/10.1103/PhysRevLett.112.236802} {\bibfield
  {journal} {\bibinfo  {journal} {Phys. Rev. Lett.}\ }\textbf {\bibinfo
  {volume} {112}},\ \bibinfo {pages} {29} (\bibinfo {year} {2014})}\BibitemShut
  {NoStop}%
\bibitem [{\citenamefont {De~Liberato}\ and\ \citenamefont
  {Ciuti}(2009)}]{de2009stimulated}%
  \BibitemOpen
  \bibfield  {author} {\bibinfo {author} {\bibfnamefont {S.}~\bibnamefont
  {De~Liberato}}\ and\ \bibinfo {author} {\bibfnamefont {C.}~\bibnamefont
  {Ciuti}},\ }\href {https://doi.org/10.1103/PhysRevLett.102.136403} {\bibfield
   {journal} {\bibinfo  {journal} {Phys. Rev. Lett.}\ }\textbf {\bibinfo
  {volume} {102}},\ \bibinfo {pages} {136403} (\bibinfo {year}
  {2009})}\BibitemShut {NoStop}%
\bibitem [{\citenamefont {Wu}\ \emph {et~al.}(2015)\citenamefont {Wu},
  \citenamefont {Buckley}, \citenamefont {Schaibley}, \citenamefont {Feng},
  \citenamefont {Yan}, \citenamefont {Mandrus}, \citenamefont {Hatami},
  \citenamefont {Yao}, \citenamefont {Vu{\v{c}}kovi{\'c}}, \citenamefont
  {Majumdar} \emph {et~al.}}]{wu2015monolayer}%
  \BibitemOpen
  \bibfield  {author} {\bibinfo {author} {\bibfnamefont {S.}~\bibnamefont
  {Wu}}, \bibinfo {author} {\bibfnamefont {S.}~\bibnamefont {Buckley}},
  \bibinfo {author} {\bibfnamefont {J.~R.}\ \bibnamefont {Schaibley}}, \bibinfo
  {author} {\bibfnamefont {L.}~\bibnamefont {Feng}}, \bibinfo {author}
  {\bibfnamefont {J.}~\bibnamefont {Yan}}, \bibinfo {author} {\bibfnamefont
  {D.~G.}\ \bibnamefont {Mandrus}}, \bibinfo {author} {\bibfnamefont
  {F.}~\bibnamefont {Hatami}}, \bibinfo {author} {\bibfnamefont
  {W.}~\bibnamefont {Yao}}, \bibinfo {author} {\bibfnamefont {J.}~\bibnamefont
  {Vu{\v{c}}kovi{\'c}}}, \bibinfo {author} {\bibfnamefont {A.}~\bibnamefont
  {Majumdar}}, \emph {et~al.},\ }\href
  {https://doi.org/https://doi.org/10.1038/nature14290} {\bibfield  {journal}
  {\bibinfo  {journal} {Nature}\ }\textbf {\bibinfo {volume} {520}},\ \bibinfo
  {pages} {69} (\bibinfo {year} {2015})}\BibitemShut {NoStop}%
\bibitem [{\citenamefont {Van~Loo}\ \emph {et~al.}(2013)\citenamefont
  {Van~Loo}, \citenamefont {Fedorov}, \citenamefont {Lalumiere}, \citenamefont
  {Sanders}, \citenamefont {Blais},\ and\ \citenamefont
  {Wallraff}}]{van2013photon}%
  \BibitemOpen
  \bibfield  {author} {\bibinfo {author} {\bibfnamefont {A.~F.}\ \bibnamefont
  {Van~Loo}}, \bibinfo {author} {\bibfnamefont {A.}~\bibnamefont {Fedorov}},
  \bibinfo {author} {\bibfnamefont {K.}~\bibnamefont {Lalumiere}}, \bibinfo
  {author} {\bibfnamefont {B.~C.}\ \bibnamefont {Sanders}}, \bibinfo {author}
  {\bibfnamefont {A.}~\bibnamefont {Blais}},\ and\ \bibinfo {author}
  {\bibfnamefont {A.}~\bibnamefont {Wallraff}},\ }\href
  {https://doi.org/10.1126/science.1244324} {\bibfield  {journal} {\bibinfo
  {journal} {Science}\ }\textbf {\bibinfo {volume} {342}},\ \bibinfo {pages}
  {1494} (\bibinfo {year} {2013})}\BibitemShut {NoStop}%
\bibitem [{\citenamefont {Mu{\~n}oz-Matutano}\ \emph
  {et~al.}(2019)\citenamefont {Mu{\~n}oz-Matutano}, \citenamefont {Wood},
  \citenamefont {Johnson}, \citenamefont {Asensio}, \citenamefont {Baragiola},
  \citenamefont {Reinhard}, \citenamefont {Lemaitre}, \citenamefont {Bloch},
  \citenamefont {Amo}, \citenamefont {Besga} \emph
  {et~al.}}]{munoz2019emergence}%
  \BibitemOpen
  \bibfield  {author} {\bibinfo {author} {\bibfnamefont {G.}~\bibnamefont
  {Mu{\~n}oz-Matutano}}, \bibinfo {author} {\bibfnamefont {A.}~\bibnamefont
  {Wood}}, \bibinfo {author} {\bibfnamefont {M.}~\bibnamefont {Johnson}},
  \bibinfo {author} {\bibfnamefont {X.~V.}\ \bibnamefont {Asensio}}, \bibinfo
  {author} {\bibfnamefont {B.}~\bibnamefont {Baragiola}}, \bibinfo {author}
  {\bibfnamefont {A.}~\bibnamefont {Reinhard}}, \bibinfo {author}
  {\bibfnamefont {A.}~\bibnamefont {Lemaitre}}, \bibinfo {author}
  {\bibfnamefont {J.}~\bibnamefont {Bloch}}, \bibinfo {author} {\bibfnamefont
  {A.}~\bibnamefont {Amo}}, \bibinfo {author} {\bibfnamefont {B.}~\bibnamefont
  {Besga}}, \emph {et~al.},\ }\href {https://doi.org/10.1038/s41563-019-0281-z}
  {\bibfield  {journal} {\bibinfo  {journal} {Nat. Mater.}\ }\textbf {\bibinfo
  {volume} {18}},\ \bibinfo {pages} {213} (\bibinfo {year} {2019})}\BibitemShut
  {NoStop}%
\bibitem [{\citenamefont {Haus}(1984)}]{haus1984waves}%
  \BibitemOpen
  \bibfield  {author} {\bibinfo {author} {\bibfnamefont {H.}~\bibnamefont
  {Haus}},\ }\bibfield  {title} {\bibinfo {title} {Waves and fields in
  optoelectronics},\ }\href@noop {} {\bibfield  {journal} {\bibinfo  {journal}
  {Prentice-Hall}\ } (\bibinfo {year} {1984})}\BibitemShut {NoStop}%
\bibitem [{\citenamefont {Landy}\ \emph {et~al.}(2008)\citenamefont {Landy},
  \citenamefont {Sajuyigbe}, \citenamefont {Mock}, \citenamefont {Smith},\ and\
  \citenamefont {Padilla}}]{PhysRevLett.100.207402}%
  \BibitemOpen
  \bibfield  {author} {\bibinfo {author} {\bibfnamefont {N.}~\bibnamefont
  {Landy}}, \bibinfo {author} {\bibfnamefont {S.}~\bibnamefont {Sajuyigbe}},
  \bibinfo {author} {\bibfnamefont {J.~J.}\ \bibnamefont {Mock}}, \bibinfo
  {author} {\bibfnamefont {D.~R.}\ \bibnamefont {Smith}},\ and\ \bibinfo
  {author} {\bibfnamefont {W.~J.}\ \bibnamefont {Padilla}},\ }\href
  {https://doi.org/10.1103/PhysRevLett.100.207402} {\bibfield  {journal}
  {\bibinfo  {journal} {Phys. Rev. Lett.}\ }\textbf {\bibinfo {volume} {100}},\
  \bibinfo {pages} {207402} (\bibinfo {year} {2008})}\BibitemShut {NoStop}%
\bibitem [{\citenamefont {Ra'di}\ \emph {et~al.}(2015)\citenamefont {Ra'di},
  \citenamefont {Simovski},\ and\ \citenamefont {Tretyakov}}]{Radi2015}%
  \BibitemOpen
  \bibfield  {author} {\bibinfo {author} {\bibfnamefont {Y.}~\bibnamefont
  {Ra'di}}, \bibinfo {author} {\bibfnamefont {C.~R.}\ \bibnamefont
  {Simovski}},\ and\ \bibinfo {author} {\bibfnamefont {S.~A.}\ \bibnamefont
  {Tretyakov}},\ }\href {https://doi.org/10.1103/PhysRevApplied.3.037001}
  {\bibfield  {journal} {\bibinfo  {journal} {Phys. Rev. Appl.}\ }\textbf
  {\bibinfo {volume} {3}},\ \bibinfo {pages} {1} (\bibinfo {year}
  {2015})}\BibitemShut {NoStop}%
\bibitem [{\citenamefont {T{\"{o}}rm{\"{o}}}\ and\ \citenamefont
  {Barnes}(2015)}]{Tormo2015}%
  \BibitemOpen
  \bibfield  {author} {\bibinfo {author} {\bibfnamefont {P.}~\bibnamefont
  {T{\"{o}}rm{\"{o}}}}\ and\ \bibinfo {author} {\bibfnamefont {W.~L.}\
  \bibnamefont {Barnes}},\ }\bibfield  {journal} {\bibinfo  {journal} {Reports
  Prog. Phys.}\ }\textbf {\bibinfo {volume} {78}},\ \href
  {https://doi.org/10.1088/0034-4885/78/1/013901}
  {10.1088/0034-4885/78/1/013901} (\bibinfo {year} {2015})\BibitemShut
  {NoStop}%
\bibitem [{\citenamefont {Piper}\ \emph {et~al.}(2014)\citenamefont {Piper},
  \citenamefont {Liu},\ and\ \citenamefont {Fan}}]{piper2014total}%
  \BibitemOpen
  \bibfield  {author} {\bibinfo {author} {\bibfnamefont {J.~R.}\ \bibnamefont
  {Piper}}, \bibinfo {author} {\bibfnamefont {V.}~\bibnamefont {Liu}},\ and\
  \bibinfo {author} {\bibfnamefont {S.}~\bibnamefont {Fan}},\ }\href
  {https://doi.org/https://doi.org/10.1063/1.4885517} {\bibfield  {journal}
  {\bibinfo  {journal} {Appl. Phys. Lett.}\ }\textbf {\bibinfo {volume}
  {104}},\ \bibinfo {pages} {251110} (\bibinfo {year} {2014})}\BibitemShut
  {NoStop}%
\bibitem [{\citenamefont {Xiao}\ \emph {et~al.}(2020)\citenamefont {Xiao},
  \citenamefont {Liu}, \citenamefont {Wang}, \citenamefont {Liu},\ and\
  \citenamefont {Zhou}}]{Xiao2020}%
  \BibitemOpen
  \bibfield  {author} {\bibinfo {author} {\bibfnamefont {S.}~\bibnamefont
  {Xiao}}, \bibinfo {author} {\bibfnamefont {T.}~\bibnamefont {Liu}}, \bibinfo
  {author} {\bibfnamefont {X.}~\bibnamefont {Wang}}, \bibinfo {author}
  {\bibfnamefont {X.}~\bibnamefont {Liu}},\ and\ \bibinfo {author}
  {\bibfnamefont {C.}~\bibnamefont {Zhou}},\ }\href
  {https://doi.org/https://doi.org/10.1103/PhysRevB.102.085410} {\bibfield
  {journal} {\bibinfo  {journal} {Phys. Rev. B}\ }\textbf {\bibinfo {volume}
  {102}},\ \bibinfo {pages} {85410} (\bibinfo {year} {2020})}\BibitemShut
  {NoStop}%
\bibitem [{\citenamefont {Gorodetsky}\ and\ \citenamefont
  {Ilchenko}(1999)}]{gorodetsky1999optical}%
  \BibitemOpen
  \bibfield  {author} {\bibinfo {author} {\bibfnamefont {M.~L.}\ \bibnamefont
  {Gorodetsky}}\ and\ \bibinfo {author} {\bibfnamefont {V.~S.}\ \bibnamefont
  {Ilchenko}},\ }\href
  {https://doi.org/https://doi.org/10.1364/JOSAB.16.000147} {\bibfield
  {journal} {\bibinfo  {journal} {JOSA B}\ }\textbf {\bibinfo {volume} {16}},\
  \bibinfo {pages} {147} (\bibinfo {year} {1999})}\BibitemShut {NoStop}%
\bibitem [{\citenamefont {Cai}\ \emph {et~al.}(2000)\citenamefont {Cai},
  \citenamefont {Painter},\ and\ \citenamefont {Vahala}}]{cai2000observation}%
  \BibitemOpen
  \bibfield  {author} {\bibinfo {author} {\bibfnamefont {M.}~\bibnamefont
  {Cai}}, \bibinfo {author} {\bibfnamefont {O.}~\bibnamefont {Painter}},\ and\
  \bibinfo {author} {\bibfnamefont {K.~J.}\ \bibnamefont {Vahala}},\ }\href
  {https://doi.org/10.1103/PhysRevLett.85.74} {\bibfield  {journal} {\bibinfo
  {journal} {Phys. Rev. Lett.}\ }\textbf {\bibinfo {volume} {85}},\ \bibinfo
  {pages} {74} (\bibinfo {year} {2000})}\BibitemShut {NoStop}%
\bibitem [{\citenamefont {Tischler}\ \emph {et~al.}(2006)\citenamefont
  {Tischler}, \citenamefont {Bradley},\ and\ \citenamefont
  {Bulovi\'{c}}}]{Tischler:06}%
  \BibitemOpen
  \bibfield  {author} {\bibinfo {author} {\bibfnamefont {J.~R.}\ \bibnamefont
  {Tischler}}, \bibinfo {author} {\bibfnamefont {M.~S.}\ \bibnamefont
  {Bradley}},\ and\ \bibinfo {author} {\bibfnamefont {V.}~\bibnamefont
  {Bulovi\'{c}}},\ }\href
  {https://doi.org/https://doi.org/10.1364/OL.31.002045} {\bibfield  {journal}
  {\bibinfo  {journal} {Opt. Lett.}\ }\textbf {\bibinfo {volume} {31}},\
  \bibinfo {pages} {2045} (\bibinfo {year} {2006})}\BibitemShut {NoStop}%
\bibitem [{\citenamefont {Noh}\ \emph {et~al.}(2012)\citenamefont {Noh},
  \citenamefont {Chong}, \citenamefont {Stone},\ and\ \citenamefont
  {Cao}}]{noh2012perfect}%
  \BibitemOpen
  \bibfield  {author} {\bibinfo {author} {\bibfnamefont {H.}~\bibnamefont
  {Noh}}, \bibinfo {author} {\bibfnamefont {Y.}~\bibnamefont {Chong}}, \bibinfo
  {author} {\bibfnamefont {A.~D.}\ \bibnamefont {Stone}},\ and\ \bibinfo
  {author} {\bibfnamefont {H.}~\bibnamefont {Cao}},\ }\href
  {https://doi.org/https://doi.org/10.1103/PhysRevLett.108.186805} {\bibfield
  {journal} {\bibinfo  {journal} {Phys. Rev. Lett.}\ }\textbf {\bibinfo
  {volume} {108}},\ \bibinfo {pages} {186805} (\bibinfo {year}
  {2012})}\BibitemShut {NoStop}%
\bibitem [{\citenamefont {Zhu}\ \emph {et~al.}(2016)\citenamefont {Zhu},
  \citenamefont {Xiao}, \citenamefont {Kang},\ and\ \citenamefont
  {Premaratne}}]{zhu2016coherent}%
  \BibitemOpen
  \bibfield  {author} {\bibinfo {author} {\bibfnamefont {W.}~\bibnamefont
  {Zhu}}, \bibinfo {author} {\bibfnamefont {F.}~\bibnamefont {Xiao}}, \bibinfo
  {author} {\bibfnamefont {M.}~\bibnamefont {Kang}},\ and\ \bibinfo {author}
  {\bibfnamefont {M.}~\bibnamefont {Premaratne}},\ }\href
  {https://doi.org/https://doi.org/10.1063/1.4944635} {\bibfield  {journal}
  {\bibinfo  {journal} {Appl. Phys. Lett}\ }\textbf {\bibinfo {volume} {108}},\
  \bibinfo {pages} {121901} (\bibinfo {year} {2016})}\BibitemShut {NoStop}%
\bibitem [{\citenamefont {Epstein}\ \emph {et~al.}(2020)\citenamefont
  {Epstein}, \citenamefont {Terr{\'e}s}, \citenamefont {Chaves}, \citenamefont
  {Pusapati}, \citenamefont {Rhodes}, \citenamefont {Frank}, \citenamefont
  {Zimmermann}, \citenamefont {Qin}, \citenamefont {Watanabe}, \citenamefont
  {Taniguchi} \emph {et~al.}}]{Epstein2020}%
  \BibitemOpen
  \bibfield  {author} {\bibinfo {author} {\bibfnamefont {I.}~\bibnamefont
  {Epstein}}, \bibinfo {author} {\bibfnamefont {B.}~\bibnamefont {Terr{\'e}s}},
  \bibinfo {author} {\bibfnamefont {A.~J.}\ \bibnamefont {Chaves}}, \bibinfo
  {author} {\bibfnamefont {V.-V.}\ \bibnamefont {Pusapati}}, \bibinfo {author}
  {\bibfnamefont {D.~A.}\ \bibnamefont {Rhodes}}, \bibinfo {author}
  {\bibfnamefont {B.}~\bibnamefont {Frank}}, \bibinfo {author} {\bibfnamefont
  {V.}~\bibnamefont {Zimmermann}}, \bibinfo {author} {\bibfnamefont
  {Y.}~\bibnamefont {Qin}}, \bibinfo {author} {\bibfnamefont {K.}~\bibnamefont
  {Watanabe}}, \bibinfo {author} {\bibfnamefont {T.}~\bibnamefont {Taniguchi}},
  \emph {et~al.},\ }\href {https://doi.org/10.1021/acs.nanolett.0c00492}
  {\bibfield  {journal} {\bibinfo  {journal} {Nano Lett.}\ }\textbf {\bibinfo
  {volume} {20}},\ \bibinfo {pages} {3545} (\bibinfo {year}
  {2020})}\BibitemShut {NoStop}%
\bibitem [{\citenamefont {Bellessa}\ \emph
  {et~al.}(2004{\natexlab{b}})\citenamefont {Bellessa}, \citenamefont
  {Bonnand}, \citenamefont {Plenet},\ and\ \citenamefont
  {Mugnier}}]{bellessa2004strong}%
  \BibitemOpen
  \bibfield  {author} {\bibinfo {author} {\bibfnamefont {J.}~\bibnamefont
  {Bellessa}}, \bibinfo {author} {\bibfnamefont {C.}~\bibnamefont {Bonnand}},
  \bibinfo {author} {\bibfnamefont {J.}~\bibnamefont {Plenet}},\ and\ \bibinfo
  {author} {\bibfnamefont {J.}~\bibnamefont {Mugnier}},\ }\href
  {https://doi.org/10.1103/PhysRevLett.93.036404} {\bibfield  {journal}
  {\bibinfo  {journal} {Phys. Rev. Lett.}\ }\textbf {\bibinfo {volume} {93}},\
  \bibinfo {pages} {036404} (\bibinfo {year} {2004}{\natexlab{b}})}\BibitemShut
  {NoStop}%
\bibitem [{\citenamefont {Liu}\ \emph {et~al.}(2014)\citenamefont {Liu},
  \citenamefont {Galfsky}, \citenamefont {Sun},\ and\ \citenamefont
  {Xia}}]{Liu2014crystal}%
  \BibitemOpen
  \bibfield  {author} {\bibinfo {author} {\bibfnamefont {X.}~\bibnamefont
  {Liu}}, \bibinfo {author} {\bibfnamefont {T.}~\bibnamefont {Galfsky}},
  \bibinfo {author} {\bibfnamefont {Z.}~\bibnamefont {Sun}},\ and\ \bibinfo
  {author} {\bibfnamefont {F.}~\bibnamefont {Xia}},\ }\href
  {https://doi.org/10.1038/nphoton.2014.304} {\bibfield  {journal} {\bibinfo
  {journal} {Nat. Photonics}\ }\textbf {\bibinfo {volume} {9}},\ \bibinfo
  {pages} {30} (\bibinfo {year} {2014})},\ \Eprint
  {https://arxiv.org/abs/1406.4826} {1406.4826} \BibitemShut {NoStop}%
\bibitem [{\citenamefont {Liu}\ \emph {et~al.}(2017)\citenamefont {Liu},
  \citenamefont {Bao}, \citenamefont {Li}, \citenamefont {Ropp}, \citenamefont
  {Wang},\ and\ \citenamefont {Zhang}}]{Liu2017}%
  \BibitemOpen
  \bibfield  {author} {\bibinfo {author} {\bibfnamefont {X.}~\bibnamefont
  {Liu}}, \bibinfo {author} {\bibfnamefont {W.}~\bibnamefont {Bao}}, \bibinfo
  {author} {\bibfnamefont {Q.}~\bibnamefont {Li}}, \bibinfo {author}
  {\bibfnamefont {C.}~\bibnamefont {Ropp}}, \bibinfo {author} {\bibfnamefont
  {Y.}~\bibnamefont {Wang}},\ and\ \bibinfo {author} {\bibfnamefont
  {X.}~\bibnamefont {Zhang}},\ }\href
  {https://doi.org/https://doi.org/10.1103/PhysRevLett.119.027403} {\bibfield
  {journal} {\bibinfo  {journal} {Phys. Rev. Lett.}\ }\textbf {\bibinfo
  {volume} {119}},\ \bibinfo {pages} {1} (\bibinfo {year} {2017})}\BibitemShut
  {NoStop}%
\bibitem [{\citenamefont {Xie}\ \emph {et~al.}(2020)\citenamefont {Xie},
  \citenamefont {Li}, \citenamefont {Chen}, \citenamefont {Chang},
  \citenamefont {Zhang}, \citenamefont {Yi},\ and\ \citenamefont
  {Wang}}]{Xie2020}%
  \BibitemOpen
  \bibfield  {author} {\bibinfo {author} {\bibfnamefont {P.}~\bibnamefont
  {Xie}}, \bibinfo {author} {\bibfnamefont {D.}~\bibnamefont {Li}}, \bibinfo
  {author} {\bibfnamefont {Y.}~\bibnamefont {Chen}}, \bibinfo {author}
  {\bibfnamefont {P.}~\bibnamefont {Chang}}, \bibinfo {author} {\bibfnamefont
  {H.}~\bibnamefont {Zhang}}, \bibinfo {author} {\bibfnamefont
  {J.}~\bibnamefont {Yi}},\ and\ \bibinfo {author} {\bibfnamefont
  {W.}~\bibnamefont {Wang}},\ }\href
  {https://doi.org/https://doi.org/10.1103/PhysRevB.102.115430} {\bibfield
  {journal} {\bibinfo  {journal} {Phys. Rev. B}\ }\textbf {\bibinfo {volume}
  {102}},\ \bibinfo {pages} {1} (\bibinfo {year} {2020})}\BibitemShut {NoStop}%
\bibitem [{\citenamefont {Cao}\ \emph {et~al.}(2020)\citenamefont {Cao},
  \citenamefont {Dong}, \citenamefont {He}, \citenamefont {Forsberg},
  \citenamefont {Jin},\ and\ \citenamefont {He}}]{cao2020normal}%
  \BibitemOpen
  \bibfield  {author} {\bibinfo {author} {\bibfnamefont {S.}~\bibnamefont
  {Cao}}, \bibinfo {author} {\bibfnamefont {H.}~\bibnamefont {Dong}}, \bibinfo
  {author} {\bibfnamefont {J.}~\bibnamefont {He}}, \bibinfo {author}
  {\bibfnamefont {E.}~\bibnamefont {Forsberg}}, \bibinfo {author}
  {\bibfnamefont {Y.}~\bibnamefont {Jin}},\ and\ \bibinfo {author}
  {\bibfnamefont {S.}~\bibnamefont {He}},\ }\href
  {https://doi.org/10.1021/acs.jpclett.0c01080} {\bibfield  {journal} {\bibinfo
   {journal} {J. Phys. Chem. Lett.}\ }\textbf {\bibinfo {volume} {11}},\
  \bibinfo {pages} {4631} (\bibinfo {year} {2020})}\BibitemShut {NoStop}%
\bibitem [{\citenamefont {Lawless}\ \emph {et~al.}(2020)\citenamefont
  {Lawless}, \citenamefont {Hrelescu}, \citenamefont {Elliott}, \citenamefont
  {Peters}, \citenamefont {McEvoy},\ and\ \citenamefont
  {Bradley}}]{lawless2020influence}%
  \BibitemOpen
  \bibfield  {author} {\bibinfo {author} {\bibfnamefont {J.}~\bibnamefont
  {Lawless}}, \bibinfo {author} {\bibfnamefont {C.}~\bibnamefont {Hrelescu}},
  \bibinfo {author} {\bibfnamefont {C.}~\bibnamefont {Elliott}}, \bibinfo
  {author} {\bibfnamefont {L.}~\bibnamefont {Peters}}, \bibinfo {author}
  {\bibfnamefont {N.}~\bibnamefont {McEvoy}},\ and\ \bibinfo {author}
  {\bibfnamefont {A.~L.}\ \bibnamefont {Bradley}},\ }\href
  {https://doi.org/10.1021/acsami.0c09261} {\bibfield  {journal} {\bibinfo
  {journal} {ACS Appl. Mater. Interfaces}\ }\textbf {\bibinfo {volume} {12}},\
  \bibinfo {pages} {46406} (\bibinfo {year} {2020})}\BibitemShut {NoStop}%
\bibitem [{\citenamefont {Simone}\ \emph {et~al.}(2014)\citenamefont {Simone},
  \citenamefont {Mezzapesa}, \citenamefont {Federica}, \citenamefont {Giorgio},
  \citenamefont {Lorenzo}, \citenamefont {Vitiello}, \citenamefont {Lucia},
  \citenamefont {Raffaele},\ and\ \citenamefont
  {Tredicucci}}]{zanotto2014perfect}%
  \BibitemOpen
  \bibfield  {author} {\bibinfo {author} {\bibfnamefont {Z.}~\bibnamefont
  {Simone}}, \bibinfo {author} {\bibfnamefont {F.~P.}\ \bibnamefont
  {Mezzapesa}}, \bibinfo {author} {\bibfnamefont {B.}~\bibnamefont {Federica}},
  \bibinfo {author} {\bibfnamefont {B.}~\bibnamefont {Giorgio}}, \bibinfo
  {author} {\bibfnamefont {B.}~\bibnamefont {Lorenzo}}, \bibinfo {author}
  {\bibfnamefont {M.~S.}\ \bibnamefont {Vitiello}}, \bibinfo {author}
  {\bibfnamefont {S.}~\bibnamefont {Lucia}}, \bibinfo {author} {\bibfnamefont
  {C.}~\bibnamefont {Raffaele}},\ and\ \bibinfo {author} {\bibfnamefont
  {A.}~\bibnamefont {Tredicucci}},\ }\href
  {https://doi.org/https://doi.org/10.1038/nphys3106} {\bibfield  {journal}
  {\bibinfo  {journal} {Nat. Phys.}\ }\textbf {\bibinfo {volume} {10}},\
  \bibinfo {pages} {830} (\bibinfo {year} {2014})}\BibitemShut {NoStop}%
\bibitem [{\citenamefont {Baldacci}\ \emph {et~al.}(2015)\citenamefont
  {Baldacci}, \citenamefont {Zanotto}, \citenamefont {Biasiol}, \citenamefont
  {Sorba},\ and\ \citenamefont {Tredicucci}}]{Baldacci2015}%
  \BibitemOpen
  \bibfield  {author} {\bibinfo {author} {\bibfnamefont {L.}~\bibnamefont
  {Baldacci}}, \bibinfo {author} {\bibfnamefont {S.}~\bibnamefont {Zanotto}},
  \bibinfo {author} {\bibfnamefont {G.}~\bibnamefont {Biasiol}}, \bibinfo
  {author} {\bibfnamefont {L.}~\bibnamefont {Sorba}},\ and\ \bibinfo {author}
  {\bibfnamefont {A.}~\bibnamefont {Tredicucci}},\ }\href
  {https://doi.org/10.1364/oe.23.009202} {\bibfield  {journal} {\bibinfo
  {journal} {Opt. Express}\ }\textbf {\bibinfo {volume} {23}},\ \bibinfo
  {pages} {9202} (\bibinfo {year} {2015})}\BibitemShut {NoStop}%
\bibitem [{\citenamefont {Li}\ \emph {et~al.}(2017)\citenamefont {Li},
  \citenamefont {Qin}, \citenamefont {Wang}, \citenamefont {Zhai},
  \citenamefont {Ren},\ and\ \citenamefont {Hu}}]{Li2017critical}%
  \BibitemOpen
  \bibfield  {author} {\bibinfo {author} {\bibfnamefont {H.}~\bibnamefont
  {Li}}, \bibinfo {author} {\bibfnamefont {M.}~\bibnamefont {Qin}}, \bibinfo
  {author} {\bibfnamefont {L.}~\bibnamefont {Wang}}, \bibinfo {author}
  {\bibfnamefont {X.}~\bibnamefont {Zhai}}, \bibinfo {author} {\bibfnamefont
  {R.}~\bibnamefont {Ren}},\ and\ \bibinfo {author} {\bibfnamefont
  {J.}~\bibnamefont {Hu}},\ }\href {https://doi.org/10.1364/oe.25.031612}
  {\bibfield  {journal} {\bibinfo  {journal} {Opt. Express}\ }\textbf {\bibinfo
  {volume} {25}},\ \bibinfo {pages} {31612} (\bibinfo {year}
  {2017})}\BibitemShut {NoStop}%
\bibitem [{\citenamefont {Qin}\ \emph {et~al.}(2021)\citenamefont {Qin},
  \citenamefont {Xiao}, \citenamefont {Liu}, \citenamefont {Ouyang},
  \citenamefont {Yu}, \citenamefont {Wang},\ and\ \citenamefont
  {Liao}}]{Qin2021}%
  \BibitemOpen
  \bibfield  {author} {\bibinfo {author} {\bibfnamefont {M.}~\bibnamefont
  {Qin}}, \bibinfo {author} {\bibfnamefont {S.}~\bibnamefont {Xiao}}, \bibinfo
  {author} {\bibfnamefont {W.}~\bibnamefont {Liu}}, \bibinfo {author}
  {\bibfnamefont {M.}~\bibnamefont {Ouyang}}, \bibinfo {author} {\bibfnamefont
  {T.}~\bibnamefont {Yu}}, \bibinfo {author} {\bibfnamefont {T.}~\bibnamefont
  {Wang}},\ and\ \bibinfo {author} {\bibfnamefont {Q.}~\bibnamefont {Liao}},\
  }\href {https://doi.org/https://doi.org/10.1364/OE.427141} {\bibfield
  {journal} {\bibinfo  {journal} {Opt. Express}\ }\textbf {\bibinfo {volume}
  {29}},\ \bibinfo {pages} {18026} (\bibinfo {year} {2021})}\BibitemShut
  {NoStop}%
\bibitem [{\citenamefont {Sanvitto}\ and\ \citenamefont
  {K{\'{e}}na-Cohen}(2016)}]{Sanvitto2016}%
  \BibitemOpen
  \bibfield  {author} {\bibinfo {author} {\bibfnamefont {D.}~\bibnamefont
  {Sanvitto}}\ and\ \bibinfo {author} {\bibfnamefont {S.}~\bibnamefont
  {K{\'{e}}na-Cohen}},\ }\href {https://doi.org/10.1038/nmat4668} {\bibfield
  {journal} {\bibinfo  {journal} {Nat. Mater.}\ }\textbf {\bibinfo {volume}
  {15}},\ \bibinfo {pages} {1061} (\bibinfo {year} {2016})}\BibitemShut
  {NoStop}%
\bibitem [{\citenamefont {Staude}\ and\ \citenamefont
  {Schilling}(2017)}]{Staude2017}%
  \BibitemOpen
  \bibfield  {author} {\bibinfo {author} {\bibfnamefont {I.}~\bibnamefont
  {Staude}}\ and\ \bibinfo {author} {\bibfnamefont {J.}~\bibnamefont
  {Schilling}},\ }\href
  {https://doi.org/https://doi.org/10.1038/nphoton.2017.39} {\bibfield
  {journal} {\bibinfo  {journal} {Nat. Photonics}\ }\textbf {\bibinfo {volume}
  {11}},\ \bibinfo {pages} {274} (\bibinfo {year} {2017})}\BibitemShut
  {NoStop}%
\bibitem [{\citenamefont {Hugall}\ \emph {et~al.}(2018)\citenamefont {Hugall},
  \citenamefont {Singh},\ and\ \citenamefont {van
  Hulst}}]{hugall2018plasmonic}%
  \BibitemOpen
  \bibfield  {author} {\bibinfo {author} {\bibfnamefont {J.~T.}\ \bibnamefont
  {Hugall}}, \bibinfo {author} {\bibfnamefont {A.}~\bibnamefont {Singh}},\ and\
  \bibinfo {author} {\bibfnamefont {N.~F.}\ \bibnamefont {van Hulst}},\ }\href
  {https://doi.org/10.1021/acsphotonics.7b01139} {\bibfield  {journal}
  {\bibinfo  {journal} {ACS Photonics}\ }\textbf {\bibinfo {volume} {5}},\
  \bibinfo {pages} {43} (\bibinfo {year} {2018})}\BibitemShut {NoStop}%
\bibitem [{\citenamefont {Koshelev}\ \emph
  {et~al.}(2018{\natexlab{a}})\citenamefont {Koshelev}, \citenamefont
  {Lepeshov}, \citenamefont {Liu}, \citenamefont {Bogdanov},\ and\
  \citenamefont {Kivshar}}]{koshelev2018asymmetric}%
  \BibitemOpen
  \bibfield  {author} {\bibinfo {author} {\bibfnamefont {K.}~\bibnamefont
  {Koshelev}}, \bibinfo {author} {\bibfnamefont {S.}~\bibnamefont {Lepeshov}},
  \bibinfo {author} {\bibfnamefont {M.}~\bibnamefont {Liu}}, \bibinfo {author}
  {\bibfnamefont {A.}~\bibnamefont {Bogdanov}},\ and\ \bibinfo {author}
  {\bibfnamefont {Y.}~\bibnamefont {Kivshar}},\ }\href
  {https://doi.org/10.1103/PhysRevLett.121.193903} {\bibfield  {journal}
  {\bibinfo  {journal} {Phys. Rev. Lett.}\ }\textbf {\bibinfo {volume} {121}},\
  \bibinfo {pages} {193903} (\bibinfo {year} {2018}{\natexlab{a}})}\BibitemShut
  {NoStop}%
\bibitem [{\citenamefont {Koshelev}\ \emph
  {et~al.}(2018{\natexlab{b}})\citenamefont {Koshelev}, \citenamefont {Sychev},
  \citenamefont {Sadrieva}, \citenamefont {Bogdanov},\ and\ \citenamefont
  {Iorsh}}]{koshelev2018strong}%
  \BibitemOpen
  \bibfield  {author} {\bibinfo {author} {\bibfnamefont {K.}~\bibnamefont
  {Koshelev}}, \bibinfo {author} {\bibfnamefont {S.}~\bibnamefont {Sychev}},
  \bibinfo {author} {\bibfnamefont {Z.~F.}\ \bibnamefont {Sadrieva}}, \bibinfo
  {author} {\bibfnamefont {A.~A.}\ \bibnamefont {Bogdanov}},\ and\ \bibinfo
  {author} {\bibfnamefont {I.}~\bibnamefont {Iorsh}},\ }\href
  {https://doi.org/https://doi.org/10.1103/PhysRevB.98.161113} {\bibfield
  {journal} {\bibinfo  {journal} {Phys. Rev. B}\ }\textbf {\bibinfo {volume}
  {98}},\ \bibinfo {pages} {161113} (\bibinfo {year}
  {2018}{\natexlab{b}})}\BibitemShut {NoStop}%
\bibitem [{\citenamefont {Wang}\ \emph {et~al.}(2020)\citenamefont {Wang},
  \citenamefont {Duan}, \citenamefont {Chen}, \citenamefont {Zhou},
  \citenamefont {Liu},\ and\ \citenamefont {Xiao}}]{wang2020controlling}%
  \BibitemOpen
  \bibfield  {author} {\bibinfo {author} {\bibfnamefont {X.}~\bibnamefont
  {Wang}}, \bibinfo {author} {\bibfnamefont {J.}~\bibnamefont {Duan}}, \bibinfo
  {author} {\bibfnamefont {W.}~\bibnamefont {Chen}}, \bibinfo {author}
  {\bibfnamefont {C.}~\bibnamefont {Zhou}}, \bibinfo {author} {\bibfnamefont
  {T.}~\bibnamefont {Liu}},\ and\ \bibinfo {author} {\bibfnamefont
  {S.}~\bibnamefont {Xiao}},\ }\href
  {https://doi.org/https://doi.org/10.1103/PhysRevB.102.155432} {\bibfield
  {journal} {\bibinfo  {journal} {Phys. Rev. B}\ }\textbf {\bibinfo {volume}
  {102}},\ \bibinfo {pages} {155432} (\bibinfo {year} {2020})}\BibitemShut
  {NoStop}%
\bibitem [{\citenamefont {Zhang}\ and\ \citenamefont
  {Bradley}(2021{\natexlab{a}})}]{zhang2021ultra}%
  \BibitemOpen
  \bibfield  {author} {\bibinfo {author} {\bibfnamefont {X.}~\bibnamefont
  {Zhang}}\ and\ \bibinfo {author} {\bibfnamefont {A.~L.}\ \bibnamefont
  {Bradley}},\ }\href@noop {} {\bibfield  {journal} {\bibinfo  {journal} {arXiv
  preprint arXiv:2104.03463}\ } (\bibinfo {year}
  {2021}{\natexlab{a}})}\BibitemShut {NoStop}%
\bibitem [{\citenamefont {Evlyukhin}\ \emph {et~al.}(2016)\citenamefont
  {Evlyukhin}, \citenamefont {Fischer}, \citenamefont {Reinhardt},\ and\
  \citenamefont {Chichkov}}]{evlyukhin2016optical}%
  \BibitemOpen
  \bibfield  {author} {\bibinfo {author} {\bibfnamefont {A.~B.}\ \bibnamefont
  {Evlyukhin}}, \bibinfo {author} {\bibfnamefont {T.}~\bibnamefont {Fischer}},
  \bibinfo {author} {\bibfnamefont {C.}~\bibnamefont {Reinhardt}},\ and\
  \bibinfo {author} {\bibfnamefont {B.~N.}\ \bibnamefont {Chichkov}},\ }\href
  {https://doi.org/10.1103/PhysRevB.94.205434} {\bibfield  {journal} {\bibinfo
  {journal} {Phys. Rev. B}\ }\textbf {\bibinfo {volume} {94}},\ \bibinfo
  {pages} {205434} (\bibinfo {year} {2016})}\BibitemShut {NoStop}%
\bibitem [{\citenamefont {Mak}\ \emph {et~al.}(2010)\citenamefont {Mak},
  \citenamefont {Lee}, \citenamefont {Hone}, \citenamefont {Shan},\ and\
  \citenamefont {Heinz}}]{PhysRevLett.105.136805}%
  \BibitemOpen
  \bibfield  {author} {\bibinfo {author} {\bibfnamefont {K.}~\bibnamefont
  {Mak}}, \bibinfo {author} {\bibfnamefont {C.}~\bibnamefont {Lee}}, \bibinfo
  {author} {\bibfnamefont {J.}~\bibnamefont {Hone}}, \bibinfo {author}
  {\bibfnamefont {J.}~\bibnamefont {Shan}},\ and\ \bibinfo {author}
  {\bibfnamefont {T.~F.}\ \bibnamefont {Heinz}},\ }\href
  {https://doi.org/https://doi.org/10.1103/PhysRevLett.105.136805} {\bibfield
  {journal} {\bibinfo  {journal} {Phys. Rev. Lett.}\ }\textbf {\bibinfo
  {volume} {105}},\ \bibinfo {pages} {136805} (\bibinfo {year}
  {2010})}\BibitemShut {NoStop}%
\bibitem [{\citenamefont {Li}\ \emph {et~al.}(2014)\citenamefont {Li},
  \citenamefont {Chernikov}, \citenamefont {Zhang}, \citenamefont {Rigosi},
  \citenamefont {Hill}, \citenamefont {Van Der~Zande}, \citenamefont {Chenet},
  \citenamefont {Shih}, \citenamefont {Hone},\ and\ \citenamefont
  {Heinz}}]{li2014measurement}%
  \BibitemOpen
  \bibfield  {author} {\bibinfo {author} {\bibfnamefont {Y.}~\bibnamefont
  {Li}}, \bibinfo {author} {\bibfnamefont {A.}~\bibnamefont {Chernikov}},
  \bibinfo {author} {\bibfnamefont {X.}~\bibnamefont {Zhang}}, \bibinfo
  {author} {\bibfnamefont {A.}~\bibnamefont {Rigosi}}, \bibinfo {author}
  {\bibfnamefont {H.~M.}\ \bibnamefont {Hill}}, \bibinfo {author}
  {\bibfnamefont {A.~M.}\ \bibnamefont {Van Der~Zande}}, \bibinfo {author}
  {\bibfnamefont {D.~A.}\ \bibnamefont {Chenet}}, \bibinfo {author}
  {\bibfnamefont {E.-M.}\ \bibnamefont {Shih}}, \bibinfo {author}
  {\bibfnamefont {J.}~\bibnamefont {Hone}},\ and\ \bibinfo {author}
  {\bibfnamefont {T.~F.}\ \bibnamefont {Heinz}},\ }\href
  {https://doi.org/10.1103/PhysRevB.90.205422} {\bibfield  {journal} {\bibinfo
  {journal} {Physical Review B}\ }\textbf {\bibinfo {volume} {90}},\ \bibinfo
  {pages} {205422} (\bibinfo {year} {2014})}\BibitemShut {NoStop}%
\bibitem [{\citenamefont {Fan}\ \emph {et~al.}(2003)\citenamefont {Fan},
  \citenamefont {Suh},\ and\ \citenamefont {Joannopoulos}}]{fan2003temporal}%
  \BibitemOpen
  \bibfield  {author} {\bibinfo {author} {\bibfnamefont {S.}~\bibnamefont
  {Fan}}, \bibinfo {author} {\bibfnamefont {W.}~\bibnamefont {Suh}},\ and\
  \bibinfo {author} {\bibfnamefont {J.~D.}\ \bibnamefont {Joannopoulos}},\
  }\href {https://doi.org/10.1364/josaa.20.000569} {\bibfield  {journal}
  {\bibinfo  {journal} {JOSA A}\ }\textbf {\bibinfo {volume} {20}},\ \bibinfo
  {pages} {569} (\bibinfo {year} {2003})}\BibitemShut {NoStop}%
\bibitem [{\citenamefont {Savona}\ \emph {et~al.}(1995)\citenamefont {Savona},
  \citenamefont {Andreani}, \citenamefont {Schwendimann},\ and\ \citenamefont
  {Quattropani}}]{savona1995quantum}%
  \BibitemOpen
  \bibfield  {author} {\bibinfo {author} {\bibfnamefont {V.}~\bibnamefont
  {Savona}}, \bibinfo {author} {\bibfnamefont {L.}~\bibnamefont {Andreani}},
  \bibinfo {author} {\bibfnamefont {P.}~\bibnamefont {Schwendimann}},\ and\
  \bibinfo {author} {\bibfnamefont {A.}~\bibnamefont {Quattropani}},\ }\href
  {https://doi.org/https://doi.org/10.1016/0038-1098(94)00865-5} {\bibfield
  {journal} {\bibinfo  {journal} {Solid State Commun.}\ }\textbf {\bibinfo
  {volume} {93}},\ \bibinfo {pages} {733} (\bibinfo {year} {1995})}\BibitemShut
  {NoStop}%
\bibitem [{\citenamefont {Deng}\ \emph {et~al.}(2010)\citenamefont {Deng},
  \citenamefont {Haug},\ and\ \citenamefont {Yamamoto}}]{deng2010exciton}%
  \BibitemOpen
  \bibfield  {author} {\bibinfo {author} {\bibfnamefont {H.}~\bibnamefont
  {Deng}}, \bibinfo {author} {\bibfnamefont {H.}~\bibnamefont {Haug}},\ and\
  \bibinfo {author} {\bibfnamefont {Y.}~\bibnamefont {Yamamoto}},\ }\href
  {https://doi.org/10.1103/RevModPhys.82.1489} {\bibfield  {journal} {\bibinfo
  {journal} {Reviews of modern physics}\ }\textbf {\bibinfo {volume} {82}},\
  \bibinfo {pages} {1489} (\bibinfo {year} {2010})}\BibitemShut {NoStop}%
\bibitem [{\citenamefont {Zhang}\ \emph {et~al.}(2018)\citenamefont {Zhang},
  \citenamefont {Gogna}, \citenamefont {Burg}, \citenamefont {Tutuc},\ and\
  \citenamefont {Deng}}]{zhang2018photonic}%
  \BibitemOpen
  \bibfield  {author} {\bibinfo {author} {\bibfnamefont {L.}~\bibnamefont
  {Zhang}}, \bibinfo {author} {\bibfnamefont {R.}~\bibnamefont {Gogna}},
  \bibinfo {author} {\bibfnamefont {W.}~\bibnamefont {Burg}}, \bibinfo {author}
  {\bibfnamefont {E.}~\bibnamefont {Tutuc}},\ and\ \bibinfo {author}
  {\bibfnamefont {H.}~\bibnamefont {Deng}},\ }\href
  {https://doi.org/https://doi.org/10.1038/s41467-018-03188-x} {\bibfield
  {journal} {\bibinfo  {journal} {Nat. Commun.}\ }\textbf {\bibinfo {volume}
  {9}},\ \bibinfo {pages} {1} (\bibinfo {year} {2018})}\BibitemShut {NoStop}%
\bibitem [{\citenamefont {Peng}\ \emph {et~al.}(2020)\citenamefont {Peng},
  \citenamefont {Wu}, \citenamefont {Li}, \citenamefont {Xu},\ and\
  \citenamefont {Li}}]{peng2020separation}%
  \BibitemOpen
  \bibfield  {author} {\bibinfo {author} {\bibfnamefont {R.}~\bibnamefont
  {Peng}}, \bibinfo {author} {\bibfnamefont {C.}~\bibnamefont {Wu}}, \bibinfo
  {author} {\bibfnamefont {H.}~\bibnamefont {Li}}, \bibinfo {author}
  {\bibfnamefont {X.}~\bibnamefont {Xu}},\ and\ \bibinfo {author}
  {\bibfnamefont {M.}~\bibnamefont {Li}},\ }\href
  {https://doi.org/https://doi.org/10.1103/PhysRevB.101.245418} {\bibfield
  {journal} {\bibinfo  {journal} {Phys. Rev. B}\ }\textbf {\bibinfo {volume}
  {101}},\ \bibinfo {pages} {245418} (\bibinfo {year} {2020})}\BibitemShut
  {NoStop}%
\bibitem [{\citenamefont {Zanotto}\ and\ \citenamefont
  {Tredicucci}(2016)}]{Zanotto2016}%
  \BibitemOpen
  \bibfield  {author} {\bibinfo {author} {\bibfnamefont {S.}~\bibnamefont
  {Zanotto}}\ and\ \bibinfo {author} {\bibfnamefont {A.}~\bibnamefont
  {Tredicucci}},\ }\href {https://doi.org/10.1038/srep24592} {\bibfield
  {journal} {\bibinfo  {journal} {Sci. Rep.}\ }\textbf {\bibinfo {volume}
  {6}},\ \bibinfo {pages} {1} (\bibinfo {year} {2016})},\ \Eprint
  {https://arxiv.org/abs/1509.00593} {1509.00593} \BibitemShut {NoStop}%
\bibitem [{\citenamefont {Gon{\c{c}}alves}\ \emph {et~al.}(2018)\citenamefont
  {Gon{\c{c}}alves}, \citenamefont {Bertelsen}, \citenamefont {Xiao},\ and\
  \citenamefont {Mortensen}}]{Goncalves2018}%
  \BibitemOpen
  \bibfield  {author} {\bibinfo {author} {\bibfnamefont {P.}~\bibnamefont
  {Gon{\c{c}}alves}}, \bibinfo {author} {\bibfnamefont {L.}~\bibnamefont
  {Bertelsen}}, \bibinfo {author} {\bibfnamefont {S.}~\bibnamefont {Xiao}},\
  and\ \bibinfo {author} {\bibfnamefont {N.~A.}\ \bibnamefont {Mortensen}},\
  }\bibfield  {journal} {\bibinfo  {journal} {Phys. Rev. B}\ }\textbf {\bibinfo
  {volume} {97}},\ \href {https://doi.org/10.1103/PhysRevB.97.041402}
  {10.1103/PhysRevB.97.041402} (\bibinfo {year} {2018})\BibitemShut {NoStop}%
\bibitem [{\citenamefont {Wang}\ \emph
  {et~al.}(2016{\natexlab{b}})\citenamefont {Wang}, \citenamefont {Li},
  \citenamefont {Chervy}, \citenamefont {Shalabney}, \citenamefont {Azzini},
  \citenamefont {Orgiu}, \citenamefont {Hutchison}, \citenamefont {Genet},
  \citenamefont {Samor{\`\i}},\ and\ \citenamefont
  {Ebbesen}}]{wang2016coherent}%
  \BibitemOpen
  \bibfield  {author} {\bibinfo {author} {\bibfnamefont {S.}~\bibnamefont
  {Wang}}, \bibinfo {author} {\bibfnamefont {S.}~\bibnamefont {Li}}, \bibinfo
  {author} {\bibfnamefont {T.}~\bibnamefont {Chervy}}, \bibinfo {author}
  {\bibfnamefont {A.}~\bibnamefont {Shalabney}}, \bibinfo {author}
  {\bibfnamefont {S.}~\bibnamefont {Azzini}}, \bibinfo {author} {\bibfnamefont
  {E.}~\bibnamefont {Orgiu}}, \bibinfo {author} {\bibfnamefont {J.~A.}\
  \bibnamefont {Hutchison}}, \bibinfo {author} {\bibfnamefont {C.}~\bibnamefont
  {Genet}}, \bibinfo {author} {\bibfnamefont {P.}~\bibnamefont {Samor{\`\i}}},\
  and\ \bibinfo {author} {\bibfnamefont {T.~W.}\ \bibnamefont {Ebbesen}},\
  }\href {https://doi.org/10.1021/acs.nanolett.6b01475} {\bibfield  {journal}
  {\bibinfo  {journal} {Nano Lett.}\ }\textbf {\bibinfo {volume} {16}},\
  \bibinfo {pages} {4368} (\bibinfo {year} {2016}{\natexlab{b}})}\BibitemShut
  {NoStop}%
\bibitem [{\citenamefont {Sarkar}\ \emph {et~al.}(2019)\citenamefont {Sarkar},
  \citenamefont {Gupta}, \citenamefont {Kumar}, \citenamefont {Schubert},
  \citenamefont {Probst}, \citenamefont {Joseph},\ and\ \citenamefont
  {K{\"{o}}nig}}]{sarkar2019hybridized}%
  \BibitemOpen
  \bibfield  {author} {\bibinfo {author} {\bibfnamefont {S.}~\bibnamefont
  {Sarkar}}, \bibinfo {author} {\bibfnamefont {V.}~\bibnamefont {Gupta}},
  \bibinfo {author} {\bibfnamefont {M.}~\bibnamefont {Kumar}}, \bibinfo
  {author} {\bibfnamefont {J.}~\bibnamefont {Schubert}}, \bibinfo {author}
  {\bibfnamefont {P.~T.}\ \bibnamefont {Probst}}, \bibinfo {author}
  {\bibfnamefont {J.}~\bibnamefont {Joseph}},\ and\ \bibinfo {author}
  {\bibfnamefont {T.~A.}\ \bibnamefont {K{\"{o}}nig}},\ }\href
  {https://doi.org/10.1021/acsami.8b20535} {\bibfield  {journal} {\bibinfo
  {journal} {ACS Appl. Mater. Interfaces}\ }\textbf {\bibinfo {volume} {11}},\
  \bibinfo {pages} {13752} (\bibinfo {year} {2019})}\BibitemShut {NoStop}%
\bibitem [{\citenamefont {Jackson}(1999)}]{jackson1999classical}%
  \BibitemOpen
  \bibfield  {author} {\bibinfo {author} {\bibfnamefont {J.~D.}\ \bibnamefont
  {Jackson}},\ }\href@noop {} {\bibinfo {title} {{Classical Electrodynamics,
  3rd. Wiley, New York}}} (\bibinfo {year} {1999})\BibitemShut {NoStop}%
\bibitem [{\citenamefont {Baffou}\ \emph {et~al.}(2009)\citenamefont {Baffou},
  \citenamefont {Quidant},\ and\ \citenamefont {Girard}}]{Baffou2009}%
  \BibitemOpen
  \bibfield  {author} {\bibinfo {author} {\bibfnamefont {G.}~\bibnamefont
  {Baffou}}, \bibinfo {author} {\bibfnamefont {R.}~\bibnamefont {Quidant}},\
  and\ \bibinfo {author} {\bibfnamefont {C.}~\bibnamefont {Girard}},\ }\href
  {https://doi.org/10.1063/1.3116645} {\bibfield  {journal} {\bibinfo
  {journal} {Appl. Phys. Lett.}\ }\textbf {\bibinfo {volume} {94}},\ \bibinfo
  {pages} {1} (\bibinfo {year} {2009})}\BibitemShut {NoStop}%
\bibitem [{\citenamefont {Karanikolas}\ \emph {et~al.}(2020)\citenamefont
  {Karanikolas}, \citenamefont {Thanopulos},\ and\ \citenamefont
  {Paspalakis}}]{Karanikolas2020}%
  \BibitemOpen
  \bibfield  {author} {\bibinfo {author} {\bibfnamefont {V.}~\bibnamefont
  {Karanikolas}}, \bibinfo {author} {\bibfnamefont {I.}~\bibnamefont
  {Thanopulos}},\ and\ \bibinfo {author} {\bibfnamefont {E.}~\bibnamefont
  {Paspalakis}},\ }\href {https://doi.org/10.1103/physrevresearch.2.033141}
  {\bibfield  {journal} {\bibinfo  {journal} {Phys. Rev. Res.}\ }\textbf
  {\bibinfo {volume} {2}},\ \bibinfo {pages} {1} (\bibinfo {year}
  {2020})}\BibitemShut {NoStop}%
\bibitem [{\citenamefont {Hsu}\ \emph {et~al.}(2019)\citenamefont {Hsu},
  \citenamefont {Frisenda}, \citenamefont {Schmidt}, \citenamefont {Arora},
  \citenamefont {de~Vasconcellos}, \citenamefont {Bratschitsch}, \citenamefont
  {van~der Zant},\ and\ \citenamefont {Castellanos-Gomez}}]{Hsu2019}%
  \BibitemOpen
  \bibfield  {author} {\bibinfo {author} {\bibfnamefont {C.}~\bibnamefont
  {Hsu}}, \bibinfo {author} {\bibfnamefont {R.}~\bibnamefont {Frisenda}},
  \bibinfo {author} {\bibfnamefont {R.}~\bibnamefont {Schmidt}}, \bibinfo
  {author} {\bibfnamefont {A.}~\bibnamefont {Arora}}, \bibinfo {author}
  {\bibfnamefont {S.~M.}\ \bibnamefont {de~Vasconcellos}}, \bibinfo {author}
  {\bibfnamefont {R.}~\bibnamefont {Bratschitsch}}, \bibinfo {author}
  {\bibfnamefont {H.~S.}\ \bibnamefont {van~der Zant}},\ and\ \bibinfo {author}
  {\bibfnamefont {A.}~\bibnamefont {Castellanos-Gomez}},\ }\bibfield  {journal}
  {\bibinfo  {journal} {Adv. Opt. Mater.}\ }\textbf {\bibinfo {volume} {7}},\
  \href {https://doi.org/10.1002/adom.201900239} {10.1002/adom.201900239}
  (\bibinfo {year} {2019})\BibitemShut {NoStop}%
\bibitem [{\citenamefont {Kang}\ \emph {et~al.}(2018)\citenamefont {Kang},
  \citenamefont {Chen}, \citenamefont {Sardar}, \citenamefont {Tordera},
  \citenamefont {Armakavicius}, \citenamefont {Darakchieva}, \citenamefont
  {Shegai},\ and\ \citenamefont {Jonsson}}]{Kang2018}%
  \BibitemOpen
  \bibfield  {author} {\bibinfo {author} {\bibfnamefont {E.~S.}\ \bibnamefont
  {Kang}}, \bibinfo {author} {\bibfnamefont {S.}~\bibnamefont {Chen}}, \bibinfo
  {author} {\bibfnamefont {S.}~\bibnamefont {Sardar}}, \bibinfo {author}
  {\bibfnamefont {D.}~\bibnamefont {Tordera}}, \bibinfo {author} {\bibfnamefont
  {N.}~\bibnamefont {Armakavicius}}, \bibinfo {author} {\bibfnamefont
  {V.}~\bibnamefont {Darakchieva}}, \bibinfo {author} {\bibfnamefont
  {T.}~\bibnamefont {Shegai}},\ and\ \bibinfo {author} {\bibfnamefont {M.~P.}\
  \bibnamefont {Jonsson}},\ }\href
  {https://doi.org/https://doi.org/10.1021/acsphotonics.8b00679} {\bibfield
  {journal} {\bibinfo  {journal} {ACS Photonics}\ }\textbf {\bibinfo {volume}
  {5}},\ \bibinfo {pages} {4046} (\bibinfo {year} {2018})}\BibitemShut
  {NoStop}%
\bibitem [{\citenamefont {Zhang}\ and\ \citenamefont
  {Bradley}(2021{\natexlab{b}})}]{PhysRevB.103.195419}%
  \BibitemOpen
  \bibfield  {author} {\bibinfo {author} {\bibfnamefont {X.}~\bibnamefont
  {Zhang}}\ and\ \bibinfo {author} {\bibfnamefont {A.~L.}\ \bibnamefont
  {Bradley}},\ }\href {https://doi.org/10.1103/PhysRevB.103.195419} {\bibfield
  {journal} {\bibinfo  {journal} {Phys. Rev. B}\ }\textbf {\bibinfo {volume}
  {103}},\ \bibinfo {pages} {195419} (\bibinfo {year}
  {2021}{\natexlab{b}})}\BibitemShut {NoStop}%
\end{thebibliography}%

\end{document}